\documentclass[a4paper,11pt,DIV=12]{scrartcl}
\pdfoutput=1

\usepackage{url}
\usepackage{booktabs}
\usepackage{array}
\usepackage{multirow}
\usepackage[normalem]{ulem}
\usepackage[ttscale=0.9]{libertine}
\usepackage{mathrsfs}
\usepackage[intlimits]{amsmath}
\usepackage{amssymb}
\usepackage{slashed}
\usepackage{braket}
\usepackage[titletoc,title]{appendix}
\usepackage[affil-it]{authblk}
\usepackage[numbers,sort&compress]{natbib}
\usepackage{graphicx}
\usepackage{cancel}
\usepackage[normalem]{ulem}
\usepackage{mathrsfs}
\usepackage{layouts}
\usepackage[protrusion=true,expansion,kerning=true,tracking=true,final]{microtype}
\usepackage[
	        colorlinks=true,
	        linkcolor=hblue,
	        citecolor=hgreen,
	        filecolor=hblue,
	        urlcolor=hred
	        ]{hyperref}
\usepackage[dvipsnames,x11names]{xcolor}
\definecolor{hgreen}{rgb}{0,.3,0}
\definecolor{hred}{rgb}{.3,0,0}
\definecolor{orange}{rgb}{1,0.5,0}
\definecolor{hblue}{rgb}{0,0,.3}
\definecolor{LightGray}{gray}{0.95}
\definecolor{gray}{gray}{0.6}
\usepackage{enumitem}
\usepackage[T1,LY1]{fontenc}
\usepackage{scrlayer-scrpage}
\usepackage{array}

\makeatletter
\DeclareOldFontCommand{\rm}{\normalfont\rmfamily}{\mathrm}
\DeclareOldFontCommand{\sf}{\normalfont\sffamily}{\mathsf}
\DeclareOldFontCommand{\tt}{\normalfont\ttfamily}{\mathtt}
\DeclareOldFontCommand{\bf}{\normalfont\bfseries}{\mathbf}
\DeclareOldFontCommand{\it}{\normalfont\itshape}{\mathit}
\DeclareOldFontCommand{\sl}{\normalfont\slshape}{\@nomath\sl}
\DeclareOldFontCommand{\sc}{\normalfont\scshape}{\@nomath\sc}

\AtBeginDocument{
\heavyrulewidth=1.0pt
\lightrulewidth=0.5pt
\cmidrulewidth=.03em
\belowrulesep=.4ex
\belowbottomsep=0pt
\aboverulesep=.4ex
\abovetopsep=0pt
\cmidrulesep=\doublerulesep
\cmidrulekern=.5em
\defaultaddspace=.5em
}

\newcolumntype{`}{!{\vrule width 0.5pt}}

\allowdisplaybreaks
\setcapindent{1em}

\setkomafont{captionlabel}{\bfseries}
\setkomafont{caption}{\itshape}
\setkomafont{titlehead}{\normalsize\sffamily}
\addtokomafont{title}{\boldmath}
\addtokomafont{section}{\boldmath}

\KOMAoptions{headinclude=false,
             footinclude=false,
             twoside=false,
             parskip=false,
             draft=false,
             abstract=true,
             numbers=noenddot,
             DIV=12}

\newcommand{\Lag}{\mathscr{L}}
\newcommand{\SES}{ {\rm SES}}

\titlehead{\hfill DO-TH 23/17}

\title{The Anatomy of $K^+\to\pi^+\nu\bar\nu$ Distributions}

\date{December 10, 2023}

\author[a]{Martin Gorbahn%
    \thanks{\texttt{mgorbahn@liverpool.ac.uk}}}
\author[a,b]{Ulserik Moldanazarova%
        \thanks{\texttt{psumolda@liverpool.ac.uk}}}
\author[c]{Kai Henryk Sieja%
    \thanks{\texttt{kai.sieja@tu-dortmund.de}}}
\author[c]{Emmanuel~Stamou%
    \thanks{\texttt{emmanuel.stamou@tu-dortmund.de}}}
\author[c]{Mustafa Tabet%
    \thanks{\texttt{mustafa.tabet@tu-dortmund.de}}}

\affil[a]{{\large Department of Mathematical Sciences, University of Liverpool, Liverpool L69 3BX, UK}}
\affil[b]{{\large Faculty of Physics and Technology, Karaganda Buketov University, 100028 Karaganda, Kazakhstan}}
\affil[c]{{\large Fakult\"at f\"ur Physik, TU Dortmund, D-44221 Dortmund, Germany}}

\begin{document}

\maketitle

\begin{abstract}
\normalsize
The excellent experimental prospects to measure the invisible mass spectrum of
the $K^+\to\pi^+\nu\bar\nu$ decay opens a new path to test 
generalised quark--neutrino interactions with flavour 
changing $s\to d$ transitions and 
as such to novel probes of Physics beyond the Standard Model. 
Such signals can be a consequence of new lepton-number violating or
lepton-number conserving interactions, with their interpretations
depending on the Majorana versus Dirac nature of the neutrinos.
Furthermore, the possible existence of new massive sterile neutrinos
can be tested via their distinctive imprints in the invariant mass spectrum.
Within the model-independent framework of the weak effective theory at 
dimension-six, we study the New Physics effects of Majorana and 
Dirac neutrinos on the differential distribution 
of $K^+\rightarrow \pi^+\nu\bar\nu$ allowing for 
lepton-number violating interactions and potential new sterile neutrinos.
We determine the current and expected future sensitivity on the 
corresponding $\Delta S=1$ neutral-current Wilson coefficients using the distribution 
measured by the NA62 collaboration and accounting for expected improvements 
based on the HIKE experiment. We present single-operator fits and also
determine correlations among different type of operators.
Even though we focus on $s\to d\nu\nu$ transitions, the operator bases
for Majorana and Dirac and the classification of
lepton-number-violating/conserving interactions is applicable also for
the study of $b\to s/d\nu\nu$ and $c\to u\nu\nu$ transitions relevant
in current phenomenology.
\end{abstract}

\tableofcontents

\setcounter{page}{1}

\section{Introduction\label{sec:introduction}}

The branching ratio of the rare $K^+ \to \pi^+ \nu \bar{\nu}$ decay is currently
being measured at $\mathcal{O}(35\%)$ accuracy at the NA62
experiment~\cite{NA62:2020zjw, NA62:2021zjw}, while improved upper limits for
the corresponding neutral decay mode, $K_L\to \pi^0\nu\bar\nu$,
are provided from the KOTO
experiment~\cite{KOTO:2018dsc, KOTO:2020prk}. Each event of the charged decay
mode has a so-called missing neutrino mass squared, which leads to a decay
spectrum that can be measured with increased experimental statistics. 
The existing measurements already play an important role in current phenomenology. 
They are sensitive to modifications of the Standard Model (SM) that are generated via
effective operators at $\mathcal{O}(100 \,\text{TeV})$ and also constrain feebly
interacting particles that can provide new final states contributing to the
invisible mass spectrum \cite{goudzovski:2022vbt,Buras_2014,Anzivino:2023bhp,He_2023}.

This exceptional sensitivity to New Physics (NP) is due to the high 
suppression of these decay rates in the SM and the precision reached in their 
theory predictions.
Both rare $K \to \pi \nu \bar{\nu} $ decay modes are flavour-changing neutral-current (FCNC)
processes, and as such loop-induced in the SM and additionally 
suppressed by products of off-diagonal Cabibbo--Kobayashi--Maskawa Matrix (CKM) elements.
The fact that the neutrinos are only
interacting with the weak sector of the SM results in a hard
Glashow--Iliopoulos--Maiani (GIM) mechanism \cite{Glashow:1970gm}, which 
highly suppresses light-quark contributions.
This generates a severe CKM suppression in the SM and facilitates a precise theory
prediction, where, to a very good approximation, only a single local
charged-current operator contributes below the charm scale. 
The evaluation of the operator matrix element and its Wilson coefficient
through an ongoing theoretical effort has lead to a theory
uncertainty of only $3\%$ and $1\%$ for the charged and neutral decay modes,
respectively~\cite{Buchalla:1998ba,Misiak:1999yg,Buras:2006gb,Mescia:2007kn,Brod:2008ss,Brod:2010hi}. 
This uncertainty includes the contribution of higher-dimensional operators
and light-quark contributions, which have been estimated in chiral
perturbation theory \cite{Isidori:2005xm} and can be evaluated on the
lattice \cite{Bai:2018hqu,Christ:2019dxu}. 
The overall uncertainty in the current SM prediction for the charged decay mode
\cite{brod2021updated} is still dominated by the CKM input parameters,
yet current input values \cite{Workman:2022ynf} result in an overall
theory uncertainty of $\simeq 6\%$.
This has to be compared with the expected uncertainty of $15\%$ and $5\%$
for the branching-ratio measurement at the end of NA62 and after the 
first phase of the proposed HIKE experiment \cite{HIKE:2023ext}, respectively.
These improved measurements, together with the ones for the $K_L$ mode~\cite{KOTO-II:2023}, 
will result in even tighter probes of potential NP
scenarios with flavour violation in the $s-d$ sector.

The present work is motivated by the fact that a more detailed measurement 
of the missing-mass spectrum will become available at the end of the 
NA62 experiment and during the run of HIKE. In anticipation of this
data, we will consider scenarios with heavy NP, including the possibility 
of additional massive sterile neutrinos, that can impact the missing-mass spectrum
of $K^+\to\pi^+\nu\bar\nu$. We discuss their effect in a mostly 
model-independent manner by considering all the relevant weak effective
theory operators up to dimension six for both Majorana and Dirac
neutrinos that generate a $s \to d \nu \bar{\nu}$ transition at tree level.
Here, non-zero contributions of (pseudo-)scalar and tensor
operators can distinctively modify the invariant mass spectrum from
the SM-like spectrum generated by vector and axial-vector operators.
These generalised neutrino interactions have recently received considerable
attention in the context of neutrino oscillations
\cite{AristizabalSierra:2018eqm,Altmannshofer:2018xyo,Falkowski:2019xoe,Bischer:2019ttk}
and correlations to charged current decays have for example been studied in
Ref.~\cite{Han:2020pff} in the context of 
the Standard Model Effective Field Theory (SMEFT).
In this work, we will study
their impact on the FCNC golden Kaon decay. The impact of sterile neutrinos on
(axial-)vector type operators has been studied in Ref.~\cite{Abada:2016plb}.
Working with the two, physically distinct scenarios of Dirac and Majorana
neutrinos allows us to incorporate mass effects of sterile neutrinos for all operators,
which would not be possible in a basis where neutrinos are represented
as Weyl fields \cite{Jenkins:2017jig}.
This distinction has the benefit of providing a transparent interpretation 
of a possible NP signal in terms of lepton-number conserving (LNC) or 
lepton-number violating (LNV) NP interactions.
Dirac and Majorana weak effective theories originate
from different UV theories, e.g., in the SMEFT only three Majorana 
neutrinos are present in the weak effective theory.
Instead three Dirac neutrinos in WET can originate from a LNC 
extension of SMEFT with at least three additional Weyl ($\nu$SMEFT).
Constraints on scalar-type SMEFT operators on the $K^+\to\pi^+\nu\bar\nu$ distributions,
but only using the total branching ratio, were also studied in 
Ref.~\cite{Deppisch:2020oyx,fridell2023probing} for the case of three massless Majorana
neutrinos.
Generalised neutrino interactions and their impact in $K\to\pi\nu\nu$ have 
been discussed in the context of the Weak Effective Theory (WET)
in Ref.~\cite{Li_2020}. The bounds are set employing the 
total branching ratio for three massless Majorana neutrinos.
However, an analysis that utilizes the measured distribution in the lab frame
and also incorporates the experimental cuts and backgrounds is hitherto missing.

With the current work we close this gap: we classify all WET operators for both
massive Majorana and Dirac neutrinos that can impact $K\to\pi\nu\nu$ at
dimension-six, we derive the differential decay rate in the lab frame of NA62,
i.e., as a function of the pion momentum and the invariant neutrino mass, and
provide for the first time a fit that  incorporate the full experimental
information currently available both for massless or massive Majorana or Dirac
neutrinos as well as additional sterile neutrinos.  We perform a full
statistical analysis of the sensitivity to probe the different NP Wilson
coefficients accounting for the single-event sensitivity and available
background information at the current and future-planned experimental setup.
The WET setup allows us to separate the discussion of LNC and LNV-type of 
interactions by imposing LNC on the Dirac basis. The resulting constraints
on dimension-six WET Wilson coefficients have a direct translation to 
SMEFT or the parameter space of concrete UV completions of the SM.
If the overarching theory is assumed to be SMEFT, which
does not contain more than the three Weyl neutrinos, neutrinos
are necessarily Majorana. In this case, scalar and tensor operators
contributing to $K\to\pi\nu\nu$ are necessarily LNV and 
thus dimension-seven. If more degrees of freedom are included ($\nu$SMEFT)
the corresponding operators can be dimension-six. Both cases match to 
the dimension-six Wilson coefficient studied in the current work.

The paper is organised as follows. In Section~\ref{sec:effective-theories-k} we
classify the weak effective theory relevant for $q\to q'\nu\nu$ FCNCs
and count the independent Wilson coefficients that contribute to flavour-changing quark
into neutrino decays. 
In the following Section~\ref{sec:Rates} we provide the relevant decay 
rates for zero neutrino masses, and explain our statistical treatment including 
the treatment of theoretical uncertainties.
In the next Section~\ref{sec:sensitivities} we provide the current and 
expected future sensitivities for the cases in which a single operator 
contributes (single-operator fits) as well as the correlated constraints on pairs 
of different operators.
After the conclusion we also provide in Appendix~\ref{app:DalitzDecayRates}
further formulae for the  double differential distributions 
$\mathrm{d}^2\mathrm{Br}/\mathrm{d}q^2\mathrm{d}k^2$.

\section{Weak effective theories for $K \to \pi\nu \bar{\nu}$\label{sec:effective-theories-k}}

The focus of this work is to study the possible 
impact of heavy New Physics (NP) scenarios in 
the distributions of $K^+ \to \pi^+\nu \bar{\nu}$. The goal is to also
include the possibility that the existence of additional heavy neutrinos could
provide further final states with missing energy. To this end, it is useful
and necessary to distinguish the conceptually different cases of having
Majorana-type versus Dirac-type neutrino masses for the three light neutrinos
or the possible extra sterile neutrinos. Both Majorana and Dirac operator
basis have a direct correspondence in terms of the basis written using chiral
(or equivalently two-component Weyl) fermion fields \cite{Jenkins:2017jig}.
However, unless the dimension-four mass terms are specified, the chiral basis
with more than three neutrino fields leaves the question of neutrino masses
unspecified: The additional right-handed neutrinos can combine with the
left-handed ones to give Dirac masses without the need for
lepton-number-violating mass terms. Alternatively, the observed neutrinos are
Majorana particles in which case their masses originate from Majorana mass
terms and the extra, sterile neutrinos have their own independent mass.

We will thus separately discuss two cases:
\begin{enumerate}[label=(\roman*)]
  \item Three or more Majorana neutrinos with dimension-six interactions with
    $s-d$ FCNC interactions that can include lepton-number violation.
  \item Dirac neutrinos with dimension-six interactions with
    $s-d$ FCNC interactions that conserve lepton-number.
\end{enumerate}
Each of these cases will have a distinct low-energy effective
theory. We thus decompose the effective Lagrangian relevant for
quark-FCNC transitions with neutrinos, i.e., 
$d_i\to d_j\nu\nu$ or $u_i\to u_j\nu\nu$, at the dimension-six level as
\begin{equation}
  \label{eq:typeEFT}
  \Lag^{\nu\text{WET}}_{q\to q'\nu\nu}\Big|_{\nu\text{-type}} = 
                                     \Lag^{(4)}_{\text{QCD}} + 
                                     \Lag^{(4)}_{\nu}\Big|_{\nu\text{-type}} + 
                                     \Lag^{(6),\text{SM}}_{q\to q'\nu\nu}  + 
                                     \Lag^{(6)}_{q\to q'\nu\nu}\Big|_{\nu\text{-type}}\,,
\end{equation}
where the superscript ``$(n)$'' indicates the mass-dimension of the operators
and $\nu\text{-type}$ labels the case of Majorana or Dirac neutrinos.
Apart from the usual dimension-four QCD Lagrangian involving quarks and gluons, 
the kinetic and mass term of the neutrinos reads
\begin{align}
  \label{eq:Lagnu}
  \Lag^{(4)}_\nu\Big|_\text{Majorana} 
  = \frac{1}{2}\sum_a \bar \nu_{Ma} (i\slashed{\partial} - M_a) \nu_{Ma}\,,&&&
  \Lag^{(4)}_\nu\Big|_\text{Dirac} 
  = \sum_a \bar \nu_{Da} (i\slashed{\partial} - M_a) \nu_{Da} \,,
\end{align}
where $a\geq 3$ runs over the different neutrino flavours, and 
$\nu_{Ma}$ and $\nu_{Da}$ denote four-component Majorana and Dirac fields, 
respectively.
Eq.~\eqref{eq:Lagnu} is already written in terms of mass-eigenstates 
for the neutrinos fields. Throughout this work we work in the mass-eigenstate 
basis for the neutrinos. This choice of basis implicitly also fixes the 
basis for the definition of the dimension-six Wilson coefficients.

In the present work we focus on $K\to\pi\nu\nu$. In this case the 
SM induces a single dimension-six operator below the 
charm mass~\cite{Buchalla:1995vs,brod2021updated}. 
For the $s\to d$ transition the
SM contribution can be written both in terms of Majorana or Dirac fields
\begin{equation}\
\label{eq:LagSM}
  \begin{split}
  \Lag^{(6),\text{SM}}_{s\to d\nu\nu} &= \sum\limits_{\substack{a,b = \left\{ 1,2,3 \right\} }}
  \left\{
    \begin{array}{c}
      C^{A,\,L, \,\text{SM}}_{ab 21}~ O^{A, \, L }_{ab 21}\\[0.5em]
      C^{V,\,LL,\,\text{SM}}_{ab 21}~ O^{V, \, LL}_{ab 21}
  \end{array}\right\}+\text{h.c.}\\
 &=
  -\frac{4G_F}{\sqrt2}\frac{\alpha}{2\pi s^2_w} \sum\limits_{\substack{\ell= \left\{ e,\mu,\tau \right\} \\ a,b = \left\{ 1,2,3 \right\} }}
  U_{\ell a}^{*} U_{\ell b} \left( \lambda_c X^\ell + \lambda_tX_t \right)\times 
  \left\{
    \begin{array}{r}
      - O^{A, \, L}_{ab 21} \\[0.5em]
       O^{V, \, LL}_{ab 21}
   \end{array}\right\}+\text{h.c.}\,,
  \end{split}
\end{equation}
where $G_F$ denotes the Fermi constant, $\alpha$ is the fine-structure constant,
and $s_w$ is the sine of the weak-mixing angle. The elements of the
quark-mixing matrix are comprised in the parameters
$\lambda_i = V^*_{is}V_{id}$, while $U_{\ell a}$ is the
Pontecorvo--Maki--Nakagawa--Sakata (PMNS) matrix.
The short-distance physics is described by the loop functions $X_t$ and $X^\ell$, 
where the dependence on the $\tau$ mass introduces a small off-diagonal neutrino coupling, 
which we ignore in the following.
The operator induced in the SM case reads
\begin{align}
\label{eq:sm-ops}
  O^{A, \, L}_{ab ij} = \frac{1}{2}(\overline{\nu}_{Ma}\gamma_\mu\gamma_5\, \, \nu_{Mb})(\overline{d}_i\,\gamma^\mu \,P_L\, {d}_j ) \,,&&&
  O^{V, \,LL}_{\,ab ij} =(\overline{\nu}_{Da}\gamma_\mu\, P_L\, \nu_{Db})(\overline{d}_i\,\gamma^\mu \,P_L\, {d}_j )\,.
\end{align}
in the Majorana and Dirac case, respectively. 
We have intentionally decided to directly separate the SM from NP contributions in 
Eq.~\eqref{eq:typeEFT} in order to present the constraints on the NP effects
more transparently.

We discuss the dimension-six operator basis for the Majorana and Dirac case
in Section~\ref{sec:MajoranaMassBasis} and Section~\ref{sec:DiracMassBasis},
respectively.
There are further possibilities, e.g., mixed Dirac--Majorana cases or Dirac masses
with lepton-number-violating dimension-six interactions, but as we shall see
the main effects can be illustrated within the two aforementioned cases.
The basis of operators can be equally well be adjusted by trivial flavour renaming
to study $b\to s/d\nu\nu$ or $c\to u\nu\nu$ transitions relevant, e.g, 
for $B\to K\nu\nu$ or rare charm FCNC with decays to neutrinos.

\subsection{Majorana-$\nu$ EFT with lepton-number violation\label{sec:MajoranaMassBasis}}
The dimension-six effective Lagrangian relevant for $d_i\rightarrow d_j \nu\nu$ transitions
for the case in which all neutrinos are Majorana particles, can be written as
\begin{equation}
\label{eq:lag-wet-majorana}
\Lag^{(6)}_{d\to d^\prime\nu\nu}\Big\vert_{\text{Majorana}} =
   \sum_{\substack{ I= \left\{ V,A \right\} \\
  \tau=\left\{L,R\right\}}} \sum_{f} C^{I,\tau}_f O^{I,\tau}_f  +
   \Biggl(
     \sum_{\substack{ I= \left\{ S,P,T \right\}}} \\
  \sum_{f} C^{I,L}_f O^{I,L}_f + \text{h.c.} \Biggr) \,,
\end{equation}
where $I$ denotes the type of interaction, $I = \{V,A,S,P,T\}$, 
which stand for $V$ for vector-, $A$ for axial-vector-, $S$ for scalar-, $P$ for pseudoscalar-,
and $T$ for tensor-type interactions. $\tau=\{L,R\}$ indicates the chirality 
of the quark currents, and $f$ comprises all neutrino and quark flavour indices, $f=\{abij\}$.
The full set of independent operators read
\begin{gather}
\label{eq:Majorana-basis}
\begin{aligned}
 O^{V, \, L}_{ab ij} &=\frac{1}{2}(\overline{\nu}_{Ma}\gamma_\mu\,  \nu_{Mb})(\overline{d}_i\,\gamma^\mu \,P_L\, {d}_j )\,,&
O^{V, \, R}_{ab ij} &=\frac{1}{2}(\overline{\nu}_{Ma}\gamma_\mu\,  \nu_{Mb})(\overline{d}_i\,\gamma^\mu \,P_R\, {d}_j )\,,\\
O^{A, \, L}_{ab ij} &=\tfrac{1}{2} (\overline{\nu}_{Ma}\gamma_\mu\gamma_5\, \, \nu_{Mb})(\overline{d}_i\,\gamma^\mu \,P_L\, {d}_j ) \,,&
O^{A, \, R}_{ab ij} &=\frac{1}{2}(\overline{\nu}_{Ma}\gamma_\mu\,\gamma_5\, \nu_{Mb})(\overline{d}_i\,\gamma^\mu \,P_R\, {d}_j )\,,\\
O^{S, \, L}_{ab ij} &=\frac{1}{2}(\overline{\nu}_{Ma} \, \nu_{Mb})(\overline{d}_i \,P_L\, {d}_j )\,,&
O^{P, \, L}_{ab ij} &=\frac{1}{2}(\overline{\nu}_{Ma} \,i\gamma_5\, \nu_{Mb})(\overline{d}_i \,P_L\, {d}_j )\,,\\
O^{T, \, L}_{ab ij} &=\frac{1}{2}(\overline{\nu}_{Ma} \,\sigma_{\mu\nu}\,\nu_{Mb})(\overline{d}_i \,\sigma^{\mu\nu}\,P_L\, {d}_j )\,,
\end{aligned}
\end{gather}
with $i,j=1,2,3$.

Apart from the quark-flavour dependence, the operators $O^{V/A, \, L/R}$
are self-adjoint after considering standard relations for four-component 
Majorana fields~\cite{Dreiner:2008tw}. This is not the case for the $O^{S/P/T, \, L}$ operators.
We ensure the hermiticity of the Lagrangian by imposing additional conditions
on the flavour structure of the Wilson coefficients.
The hermiticity condition on the Wilson coefficients $C^{V/A,\,L/R}_f$ is
\begin{equation}
  C^{V/A,\,L/R}_{abij} = \left( C^{V/A,\,L/R}_{baji} \right)^*\,.
\end{equation}
Additionally, there are the further conditions originating
from standard relations specific to four-component Majorana fields~\cite{Dreiner:2008tw}
\begin{equation}
  C^{I,\,\tau}_{\,ab ij} = \eta\, C^{I,\,\tau}_{\,ba ij}
  \quad\quad \text{where} \quad
  \eta = \begin{cases}
    +1, & \text{for} \quad I = A,\,S,\,P\\
    -1, & \text{for} \quad I = V,\, T \,.
  \end{cases} 
\end{equation}
The SM contribution to $K^+\to\pi^+\nu\nu$ ($\bar s\to \bar d\nu\nu$ transition)
in the Majorana case is 
induced by the $O^{A, \, L}_{ab ij}$
operators with  
$i=1$ (down quark) and $j=2$ (strange quark),
as discussed in Eq.~\eqref{eq:LagSM}.
Note that imposing lepton-number conservation forbids the neutrinos mass-term
but also the $O^{S/P/T, \, L}$ interactions, which are thus lepton-number violating.
 
For $s \to d \nu \nu$ transitions and the case of three Majorana neutrinos,
the number of independent interactions that are symmetric in the flavour of the 
three Majorana neutrinos are parametrised by 
$6 \cdot 6$ independent complex Wilson coefficients contained in
$C^{A, \, L/R}_{ab 21}$,  
$C^{S/P, \, L}_{ab 21}$, and 
$C^{S/P, \, L}_{ab 12}$, with $a \geq b$.
In addition, the interactions that are anti-symmetric in the neutrino flavours
are parametrised by $3 \cdot 4$ complex Wilson coefficients contained in 
$C^{V,\,L}_{ab 21}$,
$C^{V,\,R}_{ab 21}$,  
$C^{T,\,L}_{ab 21}$, and 
$C^{T,\,L}_{ab 12}$, with $a > b$.
The same number of $48$ independent complex Wilson coefficients
contributes to $b \to s \nu \nu$ decays, after the quark-flavour
indices are appropriately changed.

\subsection{Dirac-$\nu$ EFT with lepton-number conservation\label{sec:DiracMassBasis}}
For the case of Dirac neutrinos, the dimension-six effective Lagrangian 
with lepton-number conservation relevant for $d_i\rightarrow d_j \nu\bar\nu$ 
transitions can be written as
\begin{equation}
\label{eq:lag-wet-dirac}
\Lag^{(6)}_{d\to d'\nu\nu}\Big\vert_\text{Dirac}\!\!=\!\!
\sum_{\tau,\tau'=\left\{L,R\right\}} \!\!\sum_{f} C^{V,\,\tau\tau'}_f O^{V,\,\tau\tau'}_f  +
   \sum_{f} 
  \Bigl( 
    C^{S,\,LL}_f O^{S,\,LL}_f +
    C^{S,\,LR}_f O^{S,\,LR}_f +
    C^{T,\,LL}_f O^{T,\,LL}_f + \text{h.c.}\Bigr),
\end{equation}
where $\tau,\tau'=\{L,R\}$ indicate the chirality of the neutrino and/or quark currents and
$f$ comprises all neutrino and quark flavour indices, $f=\{abij\}$. The
independent, lepton-number conserving operators read
\begin{align}
\label{eq:Dirac-basis}
O^{V, \,LL}_{\,ab ij} &=(\overline{\nu}_{Da}\gamma_\mu\, P_L\, \nu_{Db})(\overline{d}_i\,\gamma^\mu \,P_L\, {d}_j )\,,&
O^{V, \,LR}_{\,ab ij} &=(\overline{\nu}_{Da}\gamma_\mu\, P_L\, \nu_{Db})(\overline{d}_i\,\gamma^\mu \,P_R\, {d}_j )\,,\\
O^{V, \,RL}_{\,ab ij} &=(\overline{\nu}_{Da}\gamma_\mu\, P_R\, \nu_{Db})(\overline{d}_i\,\gamma^\mu \,P_L\, {d}_j )\,,&
O^{V, \,RR}_{\,ab ij} &=(\overline{\nu}_{Da}\gamma_\mu\, P_R\, \nu_{Db})(\overline{d}_i\,\gamma^\mu \,P_R\, {d}_j )\,,\\
O^{S,\,LL}_{\,ab ij} &=(\overline{\nu}_{Da} \, P_L\, \nu_{Db})(\overline{d}_i \,P_L\, {d}_j )\,,&
O^{S,\,LR}_{\,ab ij} &= (\overline{\nu}_{Da} \, P_L\, \nu_{Db})(\overline{d}_i \,P_R\, {d}_j )\,,\\
O^{T,\,LL}_{\,ab ij} &= (\overline{\nu}_{Da} \,\sigma_{\mu\nu}\,P_L\,\nu_{Db})(\overline{d}_i \,\sigma^{\mu\nu}\,P_L\, {d}_j )\,.&&
\end{align}
In the Dirac case, the $O^{V, \,\tau\tau'}$ operators
are all self-adjoint if their flavour structure is neglected.
The hermiticity condition on their Wilson coefficients reads
\begin{equation}
  C^{V,\,\tau\tau'}_{abij} =  \left( C^{V,\,\tau\tau'}_{ba ji} \right)^*\,.
\end{equation}

In the Dirac basis, the SM contribution 
to $K^+\to\pi^+\bar\nu\nu$
is induced by the
operators $O^{V, \,LL}_{ab ij}$ with 
$i=1$ (down quark) and $j=2$ (strange quark)
and discussed in Eq.~\eqref{eq:LagSM}.
In the strict SM case of three massless neutrinos, neutrinos can be 
described by three left-handed fields without right-handed partners.
In this case, the scalar and tensor operators vanish by construction.
They do not, however, vanish if the small neutrino masses
are due to Dirac masses for which right-handed neutrinos are required.
In this case, scalar and tensor interactions
are possible {\itshape without the need} for lepton-number violation
in contrast to the Majorana case.

For 
$s \to d \nu \bar\nu$
transitions and the case of three Dirac neutrinos,
the interactions are parametrised by a total of $9 \cdot 10$ independent 
complex Wilson coefficients  
$C^{V, \, \tau \tau'}_{ab 21}$,
$C^{S, \, LL/LR}_{ab 21}$, 
$C^{S, \, LL/LR}_{ab 12}$, 
$C^{T, \, LL}_{ab 21}$, 
$C^{T, \, LL}_{ab 12}$, 
where $\tau,\tau' \in \left\{ L,R \right\}$. 
The same number of $90$
complex Wilson coefficients contributes to $b \to s \nu \nu$ decays,
after the quark flavour indices are appropriately changed.

\section{Rates and distributions for $K^+\to \pi^+\nu\bar\nu$ at NA62\label{sec:Rates}}
In this section, we derive the decay rates and the resulting distributions
for the decay $K^+\rightarrow\pi^+\nu\nu$ for the cases of Majorana and
Dirac neutrinos. We also discuss the treatment of theory and experimental
uncertainties used throughout the analysis.
Finally, we describe the statistical procedure that we shall use
to constrain the parameter-space of the effective theories based
on current and expected data from NA62 and HIKE, respectively.

\subsection{Decay rates\label{sec:DecayRates}}
To derive the differential decay rates for 
$K^+\rightarrow \pi^+\nu\overset{\scriptscriptstyle(-)}{\nu}$
we evaluate the matrix element
\begin{equation}
  {\cal M} = \Braket{\pi^+(p_\pi) 
    \overset{\scriptscriptstyle(-)}{\nu}_{\!\!\!a}
  \nu_b| H_{\mathrm{eff}} |K^+(p_K)} \,,
\end{equation}
where the effective Hamiltonian $H_\mathrm{eff}$ follows either from
the Lagrangian in Eq.~\eqref{eq:lag-wet-majorana} or from the one in
Eq.~\eqref{eq:lag-wet-dirac}.
To be precise, for the Dirac case we compute the amplitude for 
the transition $K^+\rightarrow \pi^+\bar\nu_a \nu_b$, while for the Majorana
case the one for $K^+\rightarrow \pi^+\nu_a \nu_b$.
We will mention differences between the Majorana and the Dirac computation when they arise.

The amplitude factorizes into a hadronic part
$\braket{\pi^+| O^{\mathrm{had}}_{sd} |K^+}$ and a
leptonic part $\braket{
  \overset{\scriptscriptstyle(-)}{\nu}_{\!\!\!a}
\nu_b| O^\mathrm{lep}_{ab} | 0}$.
The leptonic part is obtained using standard techniques; in particular we employ
Feynman rules for Majorana fields \cite{Denner:1992me} to treat the case of 
massive Majorana neutrinos.
The hadronic parts of the decay are determined by three 
form factors~\cite{FFV,FFT,Mescia:2007kn}
\begin{align}
  \Braket{\pi^+|\bar{d}\gamma^\mu s|K^+} &= (p_K + p_\pi)^\mu f^V(q^2) + \frac{m_{K}^2 - m_{\pi}^2}{q^2} (p_K - p_\pi)^\mu \left(f^S(q^2) - f^V(q^2)\right) \,, \\
  \Braket{\pi^+|\bar{d}s|K^+} &= \frac{m_{K}^2 - m_{\pi}^2}{m_s - m_d} f^S(q^2) \,, \\
  \Braket{\pi^+|\bar{d}\sigma^{\mu \nu}s|K^+} &= \left(p^{\mu}_{\pi} p^{\nu}_{K} - p^{\nu}_{\pi} p^{\mu}_{K} \right)\frac{f^T(q^2)}{m_{K} + m_{\pi}} \,,
\end{align}
with $m_p$ denoting the masses of mesons and quarks,
$p_\pi$ and $p_K$ the $\pi^+$ and $K^+$ momentum, respectively, and
$q^2\equiv(p_K-p_\pi)^2$.
To obtain the tensor form factor, $f^T(q^2)$, we use isospin symmetry to
relate it to the neutral form factor, finding $f^T(q^2) = 2f^T_{0}(q^2)$, with
$\Braket{\pi^0|\bar{s}\sigma^{\mu \nu}d|K^0} = \left(p^{\mu}_{\pi} p^{\nu}_{K} - p^{\nu}_{\pi} p^{\mu}_{K} \right)\sqrt{2}f^T_{0}(q^2)/(m_{K} + m_{\pi})$
evaluated in Ref.~\cite{FFT}.
Note, that the parity violating axial parts of the hadronic matrix elements do
not contribute to the decay due to the parity conservation of QCD.
The $q^2$-dependence of the remnant scalar form-factor functions reads~\cite{Mescia:2007kn,FFT}
\begin{align}
  \label{eqn:formfactors}
  f^V(q^2) =  f^+_0\left(1 + \lambda'_+ \frac{q^2}{m_\pi^2} + \lambda''_+ \frac{q^4}{2 m_\pi^4}\right) ,~
  f^S(q^2) =  f^+_0\left(1 + \lambda_0 \frac{q^2}{m_\pi^2}\right),~
  f^T(q^2) =  \frac{2f^T_0(0)}{1 - \lambda_T \, q^2} \,.
\end{align}
Table~\ref{tab:formfactorinput} contains the numerical input for the form factors.
For notational brevity we also define
\begin{align}
  F^S(q^2) &\equiv \frac{m_{K}^2 - m_{\pi}^2}{m_s - m_d} f^S(q^2) \,,&
  F^T(q^2) &\equiv \frac{f^T(q^2)}{m_{K} + m_{\pi}} \,.
\end{align}

\begin{table}
  \centering
  \label{tab:formfactorinput}
  \caption{Numerical values of the form factor parameters used.
    The values for the scalar and vector form factor are taken from
    Ref.~\cite{Mescia:2007kn,FFV}, while the parameters of the tensor form factor are taken from
    Ref.~\cite{FFT}.
  }
  \begin{tabular}{crl}
      \toprule
      {\bfseries Form factor} & \multicolumn{2}{l}{\bfseries Parameters}\\
      \midrule
      \multirow{2}{*}{$f_S$}   &   $f^+_0       =$&\!\!\!\!\!$0.9778  $ \\
                               &   $\lambda_0   =$&\!\!\!\!\!$(13.38 \pm 1.19) \cdot 10^{-3}    $   \\ &&\\
      \multirow{3}{*}{$f_V$}   &   $f^+_0       =$&\!\!\!\!\!$0.9778  $                    \\
                               &   $\lambda'_+  =$&\!\!\!\!\!$(24.82 \pm 1.10) \cdot 10^{-3}   $  \\
                               &   $\lambda''_+ =$&\!\!\!\!\!$(1.64 \pm 0.44) \cdot 10^{-3}  $  \\ &&\\
      \multirow{2}{*}{$f_T$}   &   $f^T_0       =$&\!\!\!\!\!$0.417(14_{\text{stat}})(5_{\text{syst}})  $    \\
                               &   $s_T         =$&\!\!\!\!\!$1.10(8_{\text{stat}})(11_{\text{syst}})\,\mathrm{GeV}^{-2}  $  \\
      \bottomrule
  \end{tabular}
\end{table}

The differential partial decay rate for the three-body decay is then given 
by~\cite{Workman:2022ynf}
\begin{equation}
  \mathrm{d}\Gamma \left(K^+ \to \pi^+
  \overset{\scriptscriptstyle(-)}{\nu}_{\!\!\!a}
  \nu_b\right)
  = \frac{(2 \pi)^4}{2 E_K} \left| \overline{{\cal M}} \right|^2
      \frac{\mathrm{d}^3\vec{p}_\pi}{(2 \pi)^3 E_\pi}
      \frac{\mathrm{d}^3\vec{p}_{\nu,a}}{(2 \pi)^3 E_{\nu,a}}
      \frac{\mathrm{d}^3\vec{p}_{\nu,b}}{(2 \pi)^3 E_{\nu,b}} \, ,
\end{equation}
where $\left| \overline{{\cal M}} \right|$ indicates that we have summed over the
polarisations of the final-state neutrinos.
Starting from this equation, one can derive the differential distribution
$\mathrm{d}^2 \Gamma_{ab}/\mathrm{d}q^2\mathrm{d}k^2$ in terms of the Lorentz invariant 
parameters $q^2 \equiv (p_{\nu,a} + p_{\nu,b})^2$ and $k^2 \equiv (p_\pi + p_{\nu,a})^2$ 
to obtain the characteristic for three-body decays Dalitz plot.
The corresponding expressions in the Majorana and Dirac case  are provided 
in Appendix~\ref{app:DalitzDecayRates}.

However, NA62 measures the differential branching ratio 
$\mathrm{d}\mathrm{Br}/\mathrm{d}|\vec{p}_\pi|$ in the lab frame. 
To arrive at this expression, we recursively
write the three-body phase-space integration into a convolution of two two-particle 
phase-space integrations. 
This introduces an additional integration over the variable
$q^2 = (p_{\nu,a} + p_{\nu,b})^2$ but allows one to trivially perform the
integration over both neutrino momenta. 
After the integration over the angular part of the pion momentum in the lab frame, we are
left with the double differential decay rate 
$\mathrm{d}^2\Gamma_{ab}/\mathrm{d}|\vec{p}_\pi|\mathrm{d}q^2$
in the lab frame.

For the case of a decay to two massless neutrinos, $\nu_a$ and $\nu_b$, 
the double differential partial decay width 
of $K^+\to\pi^+\nu_a\nu_b$
in the Majorana basis then reads
\begin{equation}
  \begin{split}
    \frac{\mathrm{d}^2\Gamma_{ab}}{\mathrm{d}|\vec{p}_\pi| \mathrm{d}q^2}\Bigg|_\text{Majorana} &=
    \frac{|\vec{p}_\pi|}{6144 \pi^3 E_K E_\pi |\vec{p}_K| \left(1 + \delta_{ab}\right)}\Theta(|\vec{p}_\pi|,q^2)\Bigl|_{m_a = m_b = 0} \Bigg{[}\\
      &+8 \lambda_{K\pi q^2} \left|f^V(q^2)\right|^2 \left(\left|C^{A,\,L,\,\text{SM}}_{ab12} + C^{A,\,L}_{ab12} + C^{A,\,R}_{ab12}\right|^2 + \left|C^{V,\,L}_{ab12} + C^{V,\,R}_{ab12}\right|^2\right) \\
      &+12 q^2 \left|F^S(q^2)\right|^2 \Big(\left|C^{P,\,L}_{ab12} + C^{P,\,L*}_{ab21}\right|^2 + \left|C^{S,\,L}_{ab12} + C^{S,\,L*}_{ab21}\right|^2 \Big)\\
    &+3 q^2 \lambda_{K\pi q^2}\left|F^T(q^2)\right|^2 \Big(\left|C^{T,\,L}_{ab12} - C^{T,\,L*}_{ab21}\right|^2\Big) \Bigg{]}
  \end{split}
  \label{eq:dGmasslessMajorana}
\end{equation}
with $\lambda_{K\pi q^2}\equiv\lambda(m_K^2,m_\pi^2,q^2)$ and
the K\"all\'en function $\lambda(x,y,z)\equiv x^2+y^2+z^2-2xy-2yz-2zx$.
The additional factor $1/(1 + \delta_{ab})$ accounts for 
the reduction of the phase space in the case of identical particles in the final state.
We stress that in Eq.~\eqref{eq:dGmasslessMajorana} we have already 
  used the hermiticity and Majorana conditions to obtain all contributions to the 
  neutrino final state $\nu_a\nu_b$. 
  Therefore, the total rate corresponds to the ``constrained'' sum of channels 
$\sum_{\substack{a,b\\a\leq b}}$ of all independent final states.
In Eq.~\eqref{eq:dGmasslessMajorana}, $\Theta(|\vec{p}_\pi|,q^2)$ is a
shorthand for the theta-functions
\begin{equation}
  \begin{split}
    \Theta(|\vec{p}_\pi|,q^2) \equiv ~&
     \theta (E_K-E_\pi)
     \theta \left(\sqrt{q^2}-m_a-m_b\right)\\
    \times ~&\theta \left(1-\frac{m_K^2+m_\pi^2-2 E_K E_\pi-q^2}{2 |\vec{p}_K| |\vec{p}_\pi| }\right)
     \theta \left(1+\frac{m_K^2+m_\pi^2-2 E_K E_\pi-q^2}{2 |\vec{p}_K| |\vec{p}_\pi| }\right)
     \,.
  \end{split}
\end{equation}
Analogously, we compute the differential partial decay width 
of $K^+\to\pi^+\bar\nu_a\nu_b$ for the massless Dirac-neutrino case
\begin{equation}
  \begin{split}
    \frac{\mathrm{d}^2\Gamma_{ab}}{\mathrm{d}|\vec{p}_\pi| \mathrm{d}q^2}\Bigg|_\text{Dirac} &=
    \frac{|\vec{p}_\pi|}{3072 \pi^3 E_K E_\pi |\vec{p}_K|}\Theta(|\vec{p}_\pi|,q^2)\Bigl|_{m_a = m_b = 0} \Bigg{[}\\
      & + 2 \lambda_{K\pi q^2} |f^V(q^2)|^2 \Bigl(\bigl|C^{V,\,LL,\,\text{SM}}_{ab12} + C^{V,\,LL}_{ab12} + C^{V,\,LR}_{ab12}\bigr|^2 + \bigl|C^{V,\,RL}_{ab12} + C^{V,\,RR}_{ab12}\bigr|^2\Bigr)
    \\
  & + 3 q^2 |F^S(q^2)|^2 \Bigl(\bigl|C^{S,\,LL}_{ab12}+C^{S,\,LR}_{ab12}\bigr|^2+\bigl|C^{S,\,LL}_{ba21}+C^{S,\,LR}_{ba21}\bigr|^2\Bigr)\\
  & +q^2\lambda_{K\pi q^2} |F^T(q^2)|^2 \Bigl(\bigl|C^{T,\,LL}_{ab12}\bigr|^2+\bigl|C^{T,\,LL}_{ba21}\bigr|^2\Bigr)
\Bigg{]}
  \end{split}
  \label{eq:dGmasslessDirac}
\end{equation}
Similarly to the Majorana case, also here we have used the hermiticity 
  conditions on the Wilson coefficients in order to include all contributions to the $\bar\nu_a \nu_b$ final state.
  However, contrarily to the Majorana case, particles are not the same as  antiparticles 
  and thus  the total rate is given by the unconstrained sum of channels $\sum_{a,b}$.

The total branching ratio for 
$K^+\to\pi^+\nu\overset{\scriptscriptstyle(-)}{\nu}$
is then the sum of the different neutrino channels
\begin{equation}
  \mathrm{Br}\left(
K^+\to\pi^+\nu\overset{\scriptscriptstyle(-)}{\nu}
  \right) = \Gamma_{\text{tot}}^{-1} \int_{\Omega(|\vec{p}_\pi|, q^2)} \mathrm{d}|\vec{p}_\pi| \mathrm{d}q^2 \overline{\sum_{a,b}}\frac{\mathrm{d}^2\Gamma_{ab}}{\mathrm{d}|\vec{p}_\pi| \mathrm{d}q^2} \, ,
\end{equation}
where $\Omega(|\vec{p}_\pi|, q^2)$ denotes the signal region of the NA62 experiment,
$\Gamma_{\text{tot}}$ the total decay width in the lab frame,
and the overline in $\overline{\sum}$ indicates that for the Majorana
case the sum is constrained, i.e., $a\leq b$, while for the Dirac case it
is unconstrained.

\subsubsection*{Uncertainties in theory predictions}
There are two main sources of uncertainties in the theory
predictions for the $K\to\pi\nu\nu$ rates and distributions:
{\itshape (i)} uncertainties associated to the SM Wilson coefficient
and {\itshape (ii)} uncertainties associated to the form-factors.

For {\itshape (i)}, i.e., SM Wilson coefficient,
we follow Ref.~\cite{brod2021updated}. We include all
state-of-the-art perturbative corrections and include in our analysis both
pure theory uncertainties, associated to residual scale dependencies, and
parametric uncertainties.
At the current level of precision, the dominant uncertainties are parametric,
mainly originating from the CKM input. All these uncertainties are
due to short-distance dynamics and thus do not depend on $q^2$.
We also note that in the SM a significant part of the CKM 
uncertainties drop out in the ratio of the charged decay mode 
and $\varepsilon_K$ \cite{Buras:2021nns}. This ratio is also 
theoretically very clean, given recent progress 
\cite{Brod:2019rzc,Brod:2021qvc,Brod:2022har,Bai:2023lkr} in the theory prediction of
$\varepsilon_K$ and has an overall uncertainty of $5\%$, which is
similar to the $6\%$ uncertainty for the SM prediction of the charged
decay branching ratio.

\begin{figure}[t]
  \centering
  \includegraphics[scale=0.45]{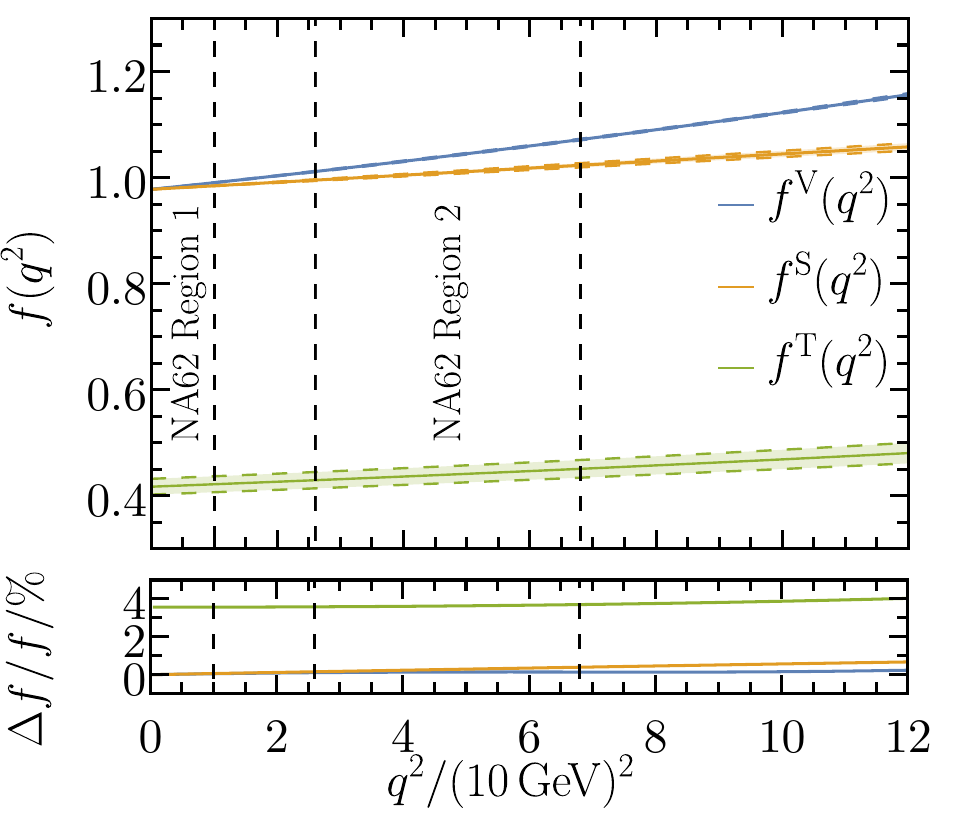}
  \caption{
    The $q^2$ dependence of the form-factor functions and their uncertainties
    as given in Table~\ref{tab:formfactorinput}.
    The coloured bands correspond to the $1\sigma$ bands for the vector, scalar and
    tensor form-factors.
    The lower panel shows the dependence of the relative uncertainty $\Delta f/f$ of each
    form-factor function.
 }
 \label{fig:FFq2dep}
\end{figure}

For {\itshape (ii)}, i.e., the form factors, we use their
numerical values and uncertainties as published in the original literature
and summarised in Table~\ref{tab:formfactorinput}.
The numerical evaluation of the form factors shows that their relative
uncertainties are roughly constant within the $q^2$ signal regions of NA62.
This is illustrated in Figure~\ref{fig:FFq2dep},
where we show the $q^2$ dependence of the three form-factor functions and their
uncertainties in the NA62 signal regions. The lower panel shows the $q^2$ dependence of
their relative uncertainties ${\Delta f}/{f}$.
Since these uncertainties are roughly flat in $q^2$, we can consider them as
global relative uncertainties, which speeds up the evaluation of the
profile likelihood ratio test. This simplifies accounting for them
in the likelihood evaluation.
Concretely, we use throughout the whole signal region an overall relative error of
$0.4\%$, $0.8\%$, and $4\%$ for the vector, scalar, and tensor form factor,
respectively, cf., Figure~\ref{fig:FFq2dep}.

The input for CKM parameters and form factors involves fits 
to experimental data. The operators considered in Eqs.~(\ref{eq:Majorana-basis}) and
(\ref{eq:Dirac-basis}) will not impact these fits, since they comprise
only a product of down-quark and neutrino field bilinears.  This can
in principle change in concrete UV completions or if the overarching
EFT is the SMEFT or $\nu$SMEFT theory. In the latter case  their Wilson coefficients 
would be correlated to other weak-effective-theory operators. 
This can result in correlations with charged-current processes 
and $\epsilon_{K}$ through weak-isospin symmetry of SMEFT and 
double insertions of SMEFT operators, respectively. 
However, the numerical impact of these
correlations is negligible, given that the
relevant SMEFT Wilson coefficients are already tightly constrained by
the currently measured $K^+\rightarrow \pi^+\nu\bar\nu$ branching
ratio. 

\subsection{Statistical treatment for EFT fit\label{sec:Statistics}}

To quantify the agreement between the experimental data and the different
NP scenarios, we perform a statistical test using a profile likelihood ratio
as test statistic that is sampled in a full frequentist manner.
The number of signal events, $s_i$, in a category $i$ is given as
\begin{align}
  \label{eq:events per bin}
  s_i = \frac{\mathrm{BR}_{K\rightarrow\pi\nu\nu}^{i}(\{C_\mathrm{NP}\})}{\SES_i} \,,
\end{align}
where $\mathrm{BR}_{K\rightarrow\pi\nu\nu}^{i}(\{C_\mathrm{NP}\})$ denotes the
branching ratio of $K^+\rightarrow\pi^+\nu\nu$ within the category $i$ for a set of
Wilson coefficients $\{C_\mathrm{NP}\}$ corresponding to the NP scenario under
consideration. Further, the variable $\SES_i$ denotes the
single-event sensitivity of category $i$ taken from Ref.~\cite{Brizioli:2021imm}.
Note that in Ref.~\cite{Brizioli:2021imm,NA62:2021zjw}, the $\SES_i$ are defined
such that  multiplied with the total branching ratio one obtains the number of expected
events in bin $i$. This is in contrast to our definition
in Eq.~\eqref{eq:events per bin} where we use the branching ratio within the corresponding
categories. We, thus, remove the phase-space information from the original definition of the $\SES_i$,
cf. Table~\ref{tab:SESeses}.

The number of expected background events, $b_i$, and observed data, $n_i$, in
each category are taken from the NA62 measurement~\cite{NA62:2021zjw}.
The events in the different categories are uncorrelated,
which implies that the total likelihood is a product of Poisson distributions
in the number of signal plus background events $s_i + b_i$.
We further neglect possible correlations between experimental parameters like
the $\SES_i$, which---if non-zero---are not provided by the NA62 collaboration.
We account for theory and experimental statistical uncertainties
in the likelihood by including the corresponding parameters as
global observables associated to auxiliary measurements.
The uncertainties of the parameters for the theory predictions described in Section~\ref{sec:DecayRates}
are all assumed to be Gaussian even though some of them, e.g., scale dependencies 
are of non-statistical nature.
However, the numerical impact of this assumption on the analysis is negligible.
The sensitivities, $\SES_i$, are constrained to follow uncorrelated Gaussian terms
while the number of expected background events follows a Poisson distribution.

\begin{table}[t]
  \centering
  \caption{The single-event sensitivities, $\SES_i$, for the respective categories
    provided by the NA62 collaboration~\cite{NA62:2019zjw,NA62:2020zjw,NA62:2021zjw,Brizioli:2021imm},
    as well as the corresponding sensitivities with the phase-space information removed.
    The definition of each category is given in the original NA62 publications.
 }
  \label{tab:SESeses}
  \begin{tabular}{l c c c c c c c c c}
    \toprule
      Category & 0 & 1 & 2 & 3 & 4 & 5 & 6 & 7 & 8 \\
    \midrule
      $\SES/10^{-10}$ from \cite{NA62:2019zjw,NA62:2020zjw,NA62:2021zjw,Brizioli:2021imm} & $3.15$ & $0.39$ & $0.54$ & $1.48$ & $0.59$ & $0.55$ & $0.63$ & $1.22$ & $1.75$ \\
      $\SES/10^{-12}$ w/o kinematic cuts & 7$6.01$ & $8.45$ & 1$6.03$ & $8.04$ & $3.20$ & $2.99$ & $3.42$ & $4.86$ & $6.97$  \\
    \bottomrule
  \end{tabular}
\end{table}

Let us now denote the total likelihood with the shorthand notation
$\mathcal{L}(x|\{C_{\mathrm{NP}}\},\nu)$, with $x$ being the outcome, i.e., the
observed data, $\{C_\mathrm{NP}\}$ the parameters of interest, i.e., the NP
Wilson coefficients, and $\nu$ the nuisance parameters.
Under the assumption that the NP scenario under consideration is realised in nature,
one can determine the confidence intervals on the model parameters $\{C_\mathrm{NP}\}$
at a given confidence level using the profile likelihood ratio $\lambda$ as
a test statistic
\begin{align}
  \lambda_{\{C_{\mathrm{NP}}\}}
  = \frac{\mathcal{L}(x|\{C_{\mathrm{NP}}\},\hat{\hat{\nu}}(\{C_{\mathrm{NP}}\}))}
  {\mathcal{L}(x|\{\hat{C}_{\mathrm{NP}}\},\hat{\nu})} \,.
\end{align}
The parameters with a single hat are the parameter values that maximise the likelihood.
The value for the nuisance parameters $\nu$ that maximise the likelihood for
given values of the parameters of interest $\{C_{\mathrm{NP}}\}$ are denoted
with a double-hat.
For given parameters of interest $\{C_{\mathrm{NP}}\}$, we calculate the corresponding
$p$-value via
\begin{align}
  p_{\{C_{\mathrm{NP}}\}} = \!\!\!\int_{\lambda^\mathrm{obs}_{\{C_{\mathrm{NP}}\}}}^{\infty}\!\!\!\!\!
  \mathrm{d}\lambda_{\{C_{\mathrm{NP}}\}} \, f(\lambda_{\{C_{\mathrm{NP}}\}} | \{C_{\mathrm{NP}}\},\nu) \,,
\end{align}
where $f$ denotes the test-statistic distribution that we sample by means of a
Monte-Carlo method under the assumption of $\{C_{\mathrm{NP}}\}$,
and with $\lambda^\mathrm{obs}_{\{C_{\mathrm{NP}}\}}$ the observed value of
the test statistic.
In order to determine the upper limit on a single Wilson coefficient $C_\mathrm{NP}$
at a confidence level of $100(1 - \alpha)\%$,
we solve for $p_{C_\mathrm{NP}} = \alpha$.
For the procedure on how to estimate two-dimensional confidence regions 
from the profile likelihood ratio see, e.g., Ref.~\cite{Rolke:2004mj}.
For further references, see Refs.~\cite{Cranmer:2014lly,Workman:2022ynf}

\section{New Physics sensitivities\label{sec:sensitivities}}

\begin{figure}[t]
  \includegraphics[width=1\textwidth]{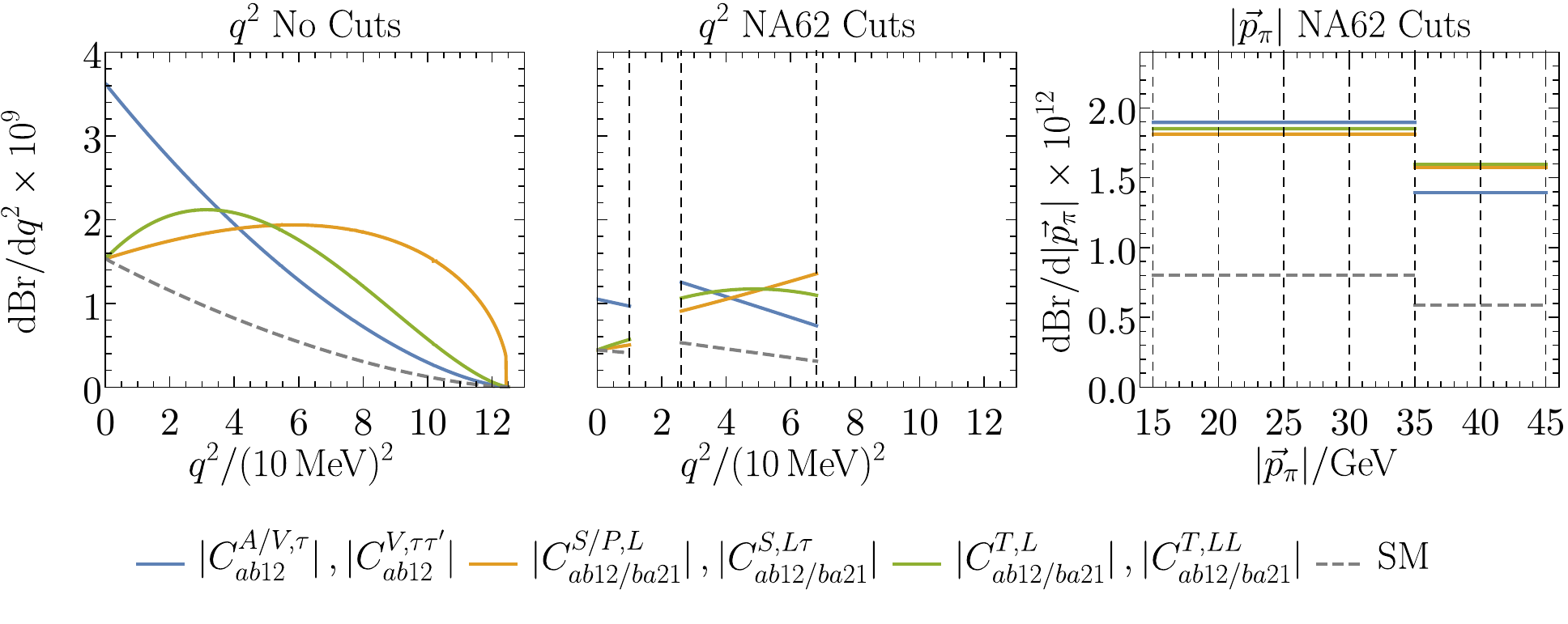}
  \caption{
    Distributions of $\mathrm{Br}(K^+ \to \pi^+ \nu \nu)$ 
    for different NP scenarios containing three massless Majorana or Dirac neutrinos.
    The left and middle panel shows differential $q^2$ distributions
    integrated over the whole physical phase space for $|\vec{p}_\pi|$
    and over $|\vec{p}_\pi|$ signal region of NA62, respectively.
    The right panel shows the lab frame $|\vec{p}_\pi|$ distribution 
    inside the NA62 $q^2$ signal regions as currently done at NA62.
    All NP Wilson coefficients that affect the rates are displayed in the legend.
    In each panel we use the values for the given Wilson coefficients that
    saturate the bound at $90\%$ CL.
    Therefore, there is no distinction between $a=b$ and $a\neq b$ cases.
    The gray, dashed line corresponds to the SM case.
    The dashed, vertical lines indicate the different NA62 signal regions.
  \label{fig:distsandNA62}}
\end{figure}

In this Section, we use the NP distributions from Section~\ref{sec:DecayRates}
together with the binned likelihood from Section~\ref{sec:Statistics}
to determine NA62's sensitivity on constraining the independent NP operators given
in Section~\ref{sec:MajoranaMassBasis} and \ref{sec:DiracMassBasis} for Majorana and
Dirac neutrinos, respectively.
We shall discuss the sensitivity based on the current experimental data of NA62~\cite{NA62:2021zjw}
and also provide estimates for the future sensitivity based on the
prospects in future experiments like HIKE.
In Section~\ref{sec:singleoperator} we switch on one NP operator at a time, while 
in Section~\ref{sec:correlations} we also show correlations by switching on pairs of
operators.
The limits on all Wilson coefficients are always
determined in addition to the Standard Model contribution, which is 
always included.

\subsection{Single-operator fits\label{sec:singleoperator}}

For the single-operator fits we discuss three different scenarios:
\begin{enumerate}[label=(\roman*)]
  \item Dirac effective theory with three light neutrinos, i.e.,
    $a,b=1,2,3$ \\
    with $m_{\nu,1}\!\approx\!m_{\nu,2}\!\approx\!m_{\nu,3}\!\approx\!0$
    (operator basis of Section~\ref{sec:DiracMassBasis}).
  \item Majorana effective theory with three light neutrinos, i.e.,
    $a,b=1,2,3$ \\
    with $m_{\nu,1}\!\approx\!m_{\nu,2}\!\approx\!m_{\nu,3}\!\approx\!0$
    (operator basis of Section~\ref{sec:MajoranaMassBasis}).
  \item Majorana effective theory with three light neutrinos
    and one additional sterile neutrino with mass $m_{\nu,4}$, i.e., $a,b=1,2,3,4$,
    (operator basis of Section~\ref{sec:MajoranaMassBasis}).
\end{enumerate}
Since in this Section we switch on one operator at a time, 
  the bounds are only sensitive to the absolute value of the respective Wilson 
  coefficients and insensitive to their CP violating phases. The exceptions are NP contributions
  that induce the same operator as the SM, i.e., $O^{A,\,L}_{aa12}$ and $O^{V,\,LL}_{aa12}$,
  for which the interference term with the SM depends on the relative phase between the
  NP and the SM Wilson coefficient. 
  Since $K^+\to\pi^+\nu\nu$ is a CP conserving decay we choose for concreteness 
  to align the NP with SM phase in the numerical analysis that follows.

\begin{table}[t]
  \centering
  \caption{
    Current and estimated future experimental lower limit at $90\%$ CL for the 
    different NP operators in the case of three massless Majorana neutrinos. 
    A single operator is turned on at a time. 
    The case labelled by $\sum a=b$ corresponds to taking
    neutrino-flavour universal Wilson coefficients, whereas in the case labelled by $a=b$ 
    a single Wilson coefficient is switched on.
    Note that in case of interference with the SM Wilson coefficient, 
    the CP phase of the NP Wilson coefficient has been aligned to the SM.
    See text for the details on the different future scenarios. 
  \label{tab:MajSensitivities}
  }
  \begin{tabular}{p{7.5em}p{4em}p{6.75em}p{6.75em}p{6.75em}}
    \toprule
    {\bfseries $\nu$-Majorana EFT} &  & {\bfseries current} & {\bfseries future}  & {\bfseries future[$q^2$ bins]}   \\
    \midrule
    \(1/\sqrt{\left|C^{V,\,L/R}_{ab12}\right|}\) 
    & $a\neq b$ & \(7.6 \cdot 10^1\,\text{TeV}\) & \(1.2 \cdot 10^2\,\text{TeV}\) & \(1.2 \cdot 10^2\,\text{TeV}\)\\
    \midrule
    \multirow{3}{*}{\(1/\sqrt{\left|C^{A,\,L/R}_{ab12}\right|}\)}
    & $\sum a = b$   & \(1.2 \cdot 10^2\,\text{TeV}\) & \(2.8 \cdot 10^2\,\text{TeV}\) & \(2.6 \cdot 10^2\,\text{TeV}\)\\
    & $a = b$   & \(8.1 \cdot 10^1\,\text{TeV}\) & \(1.7 \cdot 10^2\,\text{TeV}\) & \(1.6 \cdot 10^2\,\text{TeV}\)\\
    & $a\neq b$ & \(7.6 \cdot 10^1\,\text{TeV}\) & \(1.2 \cdot 10^2\,\text{TeV}\) & \(1.2 \cdot 10^2\,\text{TeV}\)\\
    \midrule
    \multirow{3}{*}{\(1/\sqrt{\left|C^{S,\,L}_{ab12/21}\right|}\)}
    & $\sum a = b$   & \(1.6 \cdot 10^2\,\text{TeV}\) & \(2.4 \cdot 10^2\,\text{TeV}\) & \(2.5 \cdot 10^2\,\text{TeV}\)\\
    & $a = b$   & \(1.2 \cdot 10^2\,\text{TeV}\) & \(1.9 \cdot 10^2\,\text{TeV}\) & \(1.9 \cdot 10^2\,\text{TeV}\)\\
    & $a\neq b$ & \(1.4 \cdot 10^2\,\text{TeV}\) & \(2.2 \cdot 10^2\,\text{TeV}\) & \(2.3 \cdot 10^2\,\text{TeV}\)\\
    \midrule
    \multirow{3}{*}{\(1/\sqrt{\left|C^{P,\,L}_{ab12/21}\right|}\)}
    & $\sum a = b$   & \(1.6 \cdot 10^2\,\text{TeV}\) & \(2.4 \cdot 10^2\,\text{TeV}\) & \(2.5 \cdot 10^2\,\text{TeV}\)\\
    & $a = b$   & \(1.2 \cdot 10^2\,\text{TeV}\) & \(1.9 \cdot 10^2\,\text{TeV}\) & \(1.9 \cdot 10^2\,\text{TeV}\)\\
    & $a\neq b$ & \(1.4 \cdot 10^2\,\text{TeV}\) & \(2.2 \cdot 10^2\,\text{TeV}\) & \(2.3 \cdot 10^2\,\text{TeV}\)\\
    \midrule
    \(1/\sqrt{\left|C^{T,\,L}_{ab12/21}\right|}\) 
    & $a\neq b$ & \(3.0 \cdot 10^1\,\text{TeV}\) & \(4.7 \cdot 10^1\,\text{TeV}\) & \(4.7 \cdot 10^1\,\text{TeV}\)\\
    \bottomrule
  \end{tabular}
\end{table}

\begin{table}[t]
  \centering
  \caption{
    Current and estimated future experimental lower limit at $90\%$ CL for the 
    different NP operators in the case of three massless Dirac neutrinos. 
    A single operator is turned on at a time. 
    The case labelled by $\sum a=b$ corresponds to taking
    neutrino-flavour universal Wilson coefficients, whereas in the case labelled by $a=b$ 
    a single Wilson coefficient is switched on.
    Note that in case of interference
    with the SM Wilson coefficient, the CP phase of the NP Wilson coefficient
    has been aligned to the SM.
    See text for the details on the different future scenarios.
  \label{tab:DiracSensitivities}}
  \begin{tabular}{p{7.5em}p{4em}p{6.75em}p{6.75em}p{6.75em}}
    \toprule
    {\bfseries $\nu$-Dirac EFT} &  & {\bfseries current} & {\bfseries future}  & {\bfseries future[$q^2$ bins]}   \\
    \midrule
    \multirow{3}{*}{\(1/\sqrt{\left|C^{V,L\tau}_{ab12}\right|}\)} 
    & $\sum a = b$   & \(1.2 \cdot 10^2\,\text{TeV}\) & \(2.8 \cdot 10^2\,\text{TeV}\) & \(2.6 \cdot 10^2\,\text{TeV}\)\\
    & $a = b$   & \(8.2 \cdot 10^1\,\text{TeV}\) & \(1.7 \cdot 10^2\,\text{TeV}\) & \(1.6 \cdot 10^2\,\text{TeV}\)\\
    & $a\neq b$ & \(6.3 \cdot 10^1\,\text{TeV}\) & \(1.0 \cdot 10^2\,\text{TeV}\) & \(9.6 \cdot 10^1\,\text{TeV}\)\\
    \midrule
    \multirow{3}{*}{\(1/\sqrt{\left|C^{V,R\tau}_{ab12}\right|}\)} 
    & $\sum a = b$   & \(8.4 \cdot 10^1\,\text{TeV}\) & \(1.3\cdot 10^2\,\text{TeV}\) & \(1.3 \cdot 10^2\,\text{TeV}\)\\
    & $a = b$   & \(6.3 \cdot 10^1\,\text{TeV}\) & \(1.0 \cdot 10^2\,\text{TeV}\) & \(9.6 \cdot 10^1\,\text{TeV}\)\\
    & $a\neq b$ & \(6.3 \cdot 10^1\,\text{TeV}\) & \(1.0 \cdot 10^2\,\text{TeV}\) & \(9.6 \cdot 10^1\,\text{TeV}\)\\
    \midrule
    \multirow{3}{*}{\(1/\sqrt{\left|C^{S,L\tau}_{ab 12/21}\right|}\)} 
    & $\sum a = b$   & \(1.6 \cdot 10^2\,\text{TeV}\) & \(2.5 \cdot 10^2\,\text{TeV}\) & \(2.5 \cdot 10^2\,\text{TeV}\)\\
    & $a = b$   & \(1.2 \cdot 10^2\,\text{TeV}\) & \(1.9 \cdot 10^2\,\text{TeV}\) & \(1.9 \cdot 10^2\,\text{TeV}\)\\
    & $a\neq b$ & \(1.2 \cdot 10^2\,\text{TeV}\) & \(1.9 \cdot 10^2\,\text{TeV}\) & \(1.9 \cdot 10^2\,\text{TeV}\)\\
    \midrule
    \multirow{3}{*}{\(1/\sqrt{\left|C^{T,LL}_{ab12/21}\right|}\)}
    & $\sum a = b$   & \(3.5 \cdot 10^1\,\text{TeV}\) & \(5.7 \cdot 10^1\,\text{TeV}\) & \(5.7 \cdot 10^1\,\text{TeV}\)\\
    & $a = b$   & \(2.7 \cdot 10^1\,\text{TeV}\) & \(4.2 \cdot 10^1\,\text{TeV}\) & \(4.2 \cdot 10^1\,\text{TeV}\)\\
    & $a\neq b$ & \(2.7 \cdot 10^1\,\text{TeV}\) & \(4.2 \cdot 10^1\,\text{TeV}\) & \(4.2 \cdot 10^1\,\text{TeV}\)\\
    \bottomrule
  \end{tabular}
\end{table}

Even though the cases (i) and (ii) are physically distinguishable due
to the different number of degrees of freedom there will always be a scenario
in (i) that leads to the same NP distribution as a scenario in (ii) in the limit of
$m_\nu\to 0$ if a single NP operator is switched on.
This is best seen by comparing the Majorana to the Dirac
distributions for massless neutrinos in Eqs.~\eqref{eq:dGmasslessMajorana}
and \eqref{eq:dGmasslessDirac} and is illustrated also in Figure~\ref{fig:distsandNA62},
where we present all the independent differential distributions possible for the case (i) and (ii).
Throughout Figure~\ref{fig:distsandNA62}, the value of the Wilson coefficients
is chosen to saturate the current experimental limit at $90\%$ Confidence Level (CL).
The limits are determined from the prescription of Section~\ref{sec:Statistics}.

\begin{figure}[t]
  \includegraphics[width=1\textwidth]{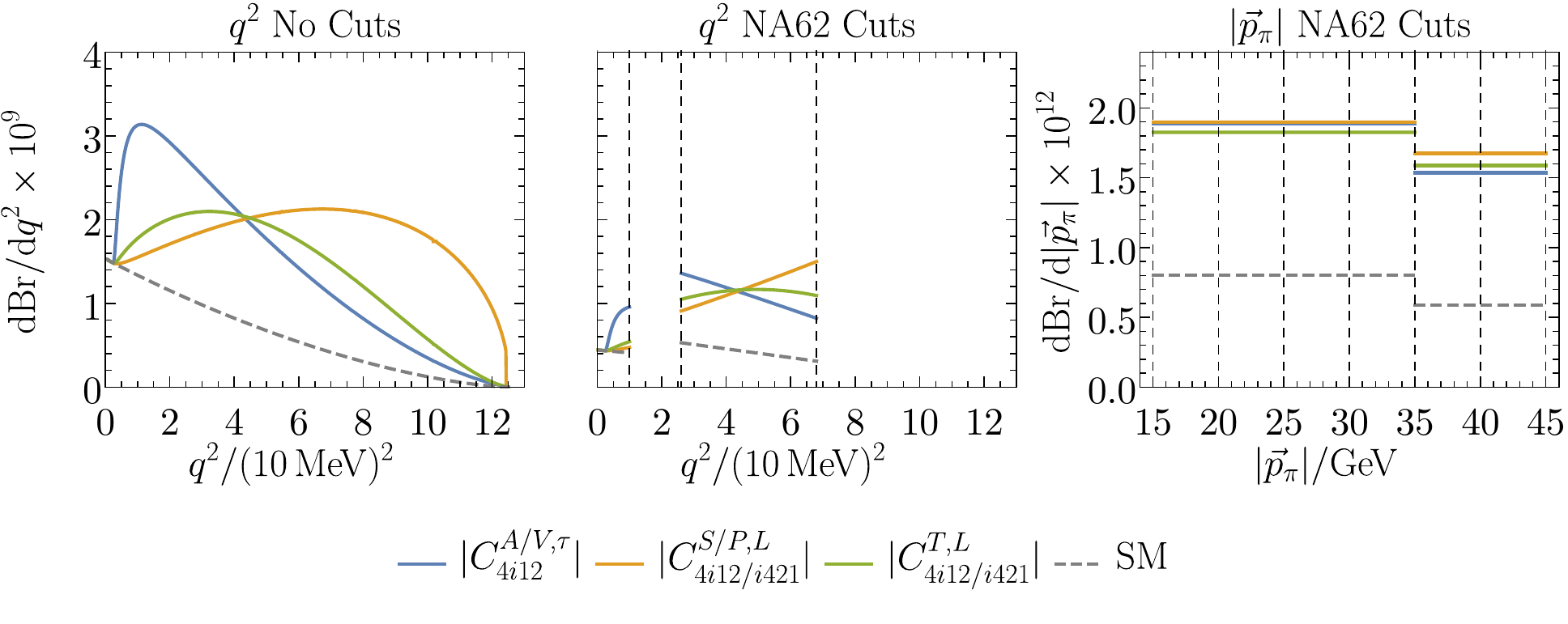}
  \caption{
    Distributions of $\mathrm{Br}(K^+ \to \pi^+ \nu \nu)$ 
    for different NP scenarios containing an additional fourth 
    massive Majorana neutrino with mass, $m_{\nu , 4} = 50 \, \text{MeV}$.
    Similarly to Figure~\ref{fig:distsandNA62},
    the left and middle panel shows differential $q^2$ distributions
    integrated over the whole physical phase space for $|\vec{p}_\pi|$
    and over the $|\vec{p}_\pi|$ signal region of NA62, respectively.
    The right panel shows the lab frame $|\vec{p}_\pi|$ distribution 
    inside the NA62 $q^2$ signal regions as currently done at NA62.
    All NP Wilson coefficients that affect the rates are displayed in the legend.
    In each panel we use the values for the given Wilson coefficients that
    saturate the bound at $90\%$ CL.
    The gray, dashed line corresponds to the SM case.
    The dashed, vertical lines indicate the different NA62 signal regions.
    \label{fig:MassiveDist}
    }
\end{figure}

In the left panel and middle panel of Figure~\ref{fig:distsandNA62}, we show
the different $\mathrm{dBr}(K^+ \rightarrow \pi^+\nu\nu)/\mathrm{d}q^2$
distributions. The difference between left and middle panel is that the former
shows the distributions integrated over the whole pion-momentum, $|\vec{p}_\pi|$, phase
space while in the latter we have integrated over the pion-momentum signal
region of NA62.  A comparison of the different shapes shows how a differential
measurement can be used to distinguish different operators.
The right panel shows the distributions
$\mathrm{dBr}(K^+ \rightarrow \pi^+\nu\nu)/\mathrm{d}|\vec{p}_\pi|$ 
in the NA62 lab frame after integrating over the $q^2$ signal region of NA62.
This corresponds exactly to the categories 3--8 currently employed by
NA62~\cite{NA62:2021zjw}. This plot also makes it evident that a binning in
the missing-momentum square, $q^2$, is able to distinguish between the different NP scenarios
within the Majorana or the Dirac basis but not between the Majorana or Dirac
nature of the neutrinos.
In Table~\ref{tab:MajSensitivities} and \ref{tab:DiracSensitivities} we collect
the constraints at $90\%$ CL on the different Wilson coefficients for 
the case of three Majorana neutrinos, i.e., scenario (i), and three Dirac neutrinos, i.e, scenario (ii), respectively.

Additionally, we perform first estimates of the future sensitivity reach 
based on the proposed HIKE experiment.
To this end we use the expected improvements highlighted
in the letter of intent for HIKE~\cite{HIKE:2022qra,HIKE:2023ext}
and perform a naive upscaling of the luminosity such that the
expected number of signal events reach 400--500 while also 
reducing the number of background events to ensure that
the intended signal to background ratio of $B/S\sim 0.5\text{--}0.7$
is achieved.\footnote{Ref.~\cite{HIKE:2023ext} and private communication with NA62.} 
Apart from this, we assume the same experimental setup as NA62. 
We take the expected data to be the sum of the expected SM
contribution plus the upscaled background of NA62 after the aforementioned 
reduction. The resulting limits are shown in the column labeled ``future'' in 
Table~\ref{tab:MajSensitivities} and \ref{tab:DiracSensitivities}
for the Majorana and Dirac case, respectively. 

For the NP scenarios (i) and (ii), we also estimate the expected limits on the
Wilson coefficients by choosing an alternative categorisation of the events.
For this we integrate over pion-momentum signal region,
but choose to bin the data in the current low-$q^2$ and 
high-$q^2$ signal region NA62 in two and four equidistant $q^2$-bins, respectively.
To achieve this, we rescale the $\SES$ of the \texttt{2018-NEWCOL} sample (category 3-8)
under the assumption that the efficiencies entering the $\SES$ 
and the background distributions are flat in $q^2$. 
The expected data is taken to be the sum of the
expected SM contribution plus the upscaled background. The estimated
sensitivities to the NP operators are shown in the right column of
Table~\ref{tab:MajSensitivities} and \ref{tab:DiracSensitivities}
for the Majorana and Dirac case, respectively. 
In both future scenarios, we observe an increase of approximately
$50\%$--$60\%$ for all NP operators apart from the diagonal vector ones
that interfere with the Standard Model. For them we observe an increase
in sensitivity of $100\%$--$130\%$.

\begin{figure}[t]
  \centering
  \includegraphics[scale=0.45]{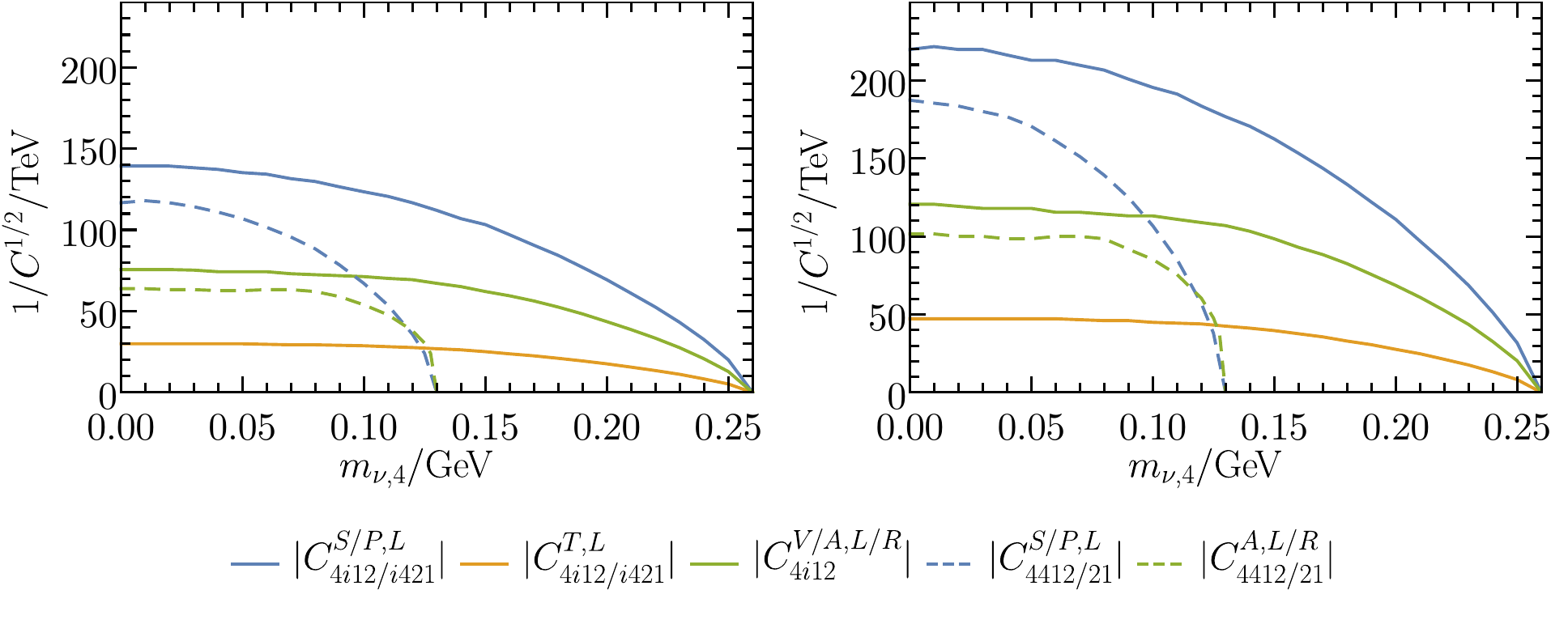}
  \caption{
    Current (left) and future (right) lower limits of $1/\sqrt{C}$ at $90\%$ CL
    for the scenarios with an additional massive fourth neutrino 
    interacting solely via one of the given NP Wilson coefficients
    as a function of its mass $m_{\nu,4}$.
  \label{fig:Covermass}}
\end{figure}

Finally, in the context of single-operator fits we also study scenario (iii).
Namely, the case of having a fourth massive Majorana neutrino, $\nu_4$,
with a mass such that the decays $K^+\to\pi^+\nu\nu_4$ or 
$K^+\to\pi^+\nu_4\nu_4$ are kinematically possible.
To illustrate the possible impact of a fourth massive neutrino on the distributions
we choose the benchmark mass $m_{\nu,4}=50$\,MeV and show in 
Figure~\ref{fig:MassiveDist} the analogous distributions as for the 
three massless neutrino cases.

Figure~\ref{fig:Covermass} shows the current constraints 
at $90\%$ CL as a function of the fourth neutrino mass $m_{\nu,4}$.
The left panel shows the current sensitivity while the right one the 
projected sensitivity.
All limits are given at $90\%$ CL. Here we observe an increase of 
$50\%$--$60\%$ in the sensitivities depending on the NP operator.
Note that the kinematic endpoint of the contribution of the diagonal 
operators, i.e., $O^{S/P,\,L}_{4412/21}$ and $O^{A,\,L/R}_{4412}$ (dashed lines), 
lies at around $\sim 130\,\text{MeV}$ as the final state necessarily 
contains two massive neutrinos.

%
%

\begin{figure}[t]
  \centering
  {\large\bfseries Massless Dirac Neutrinos}\\
  \includegraphics[scale=0.47]{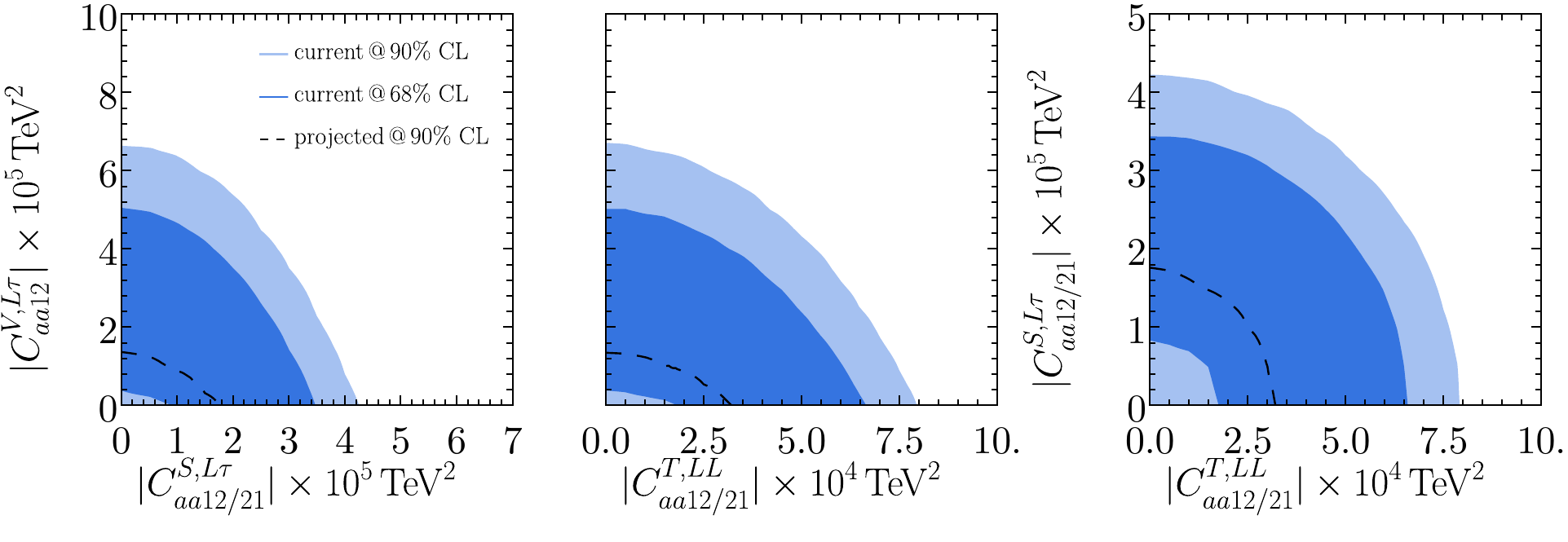}
  \includegraphics[scale=0.47]{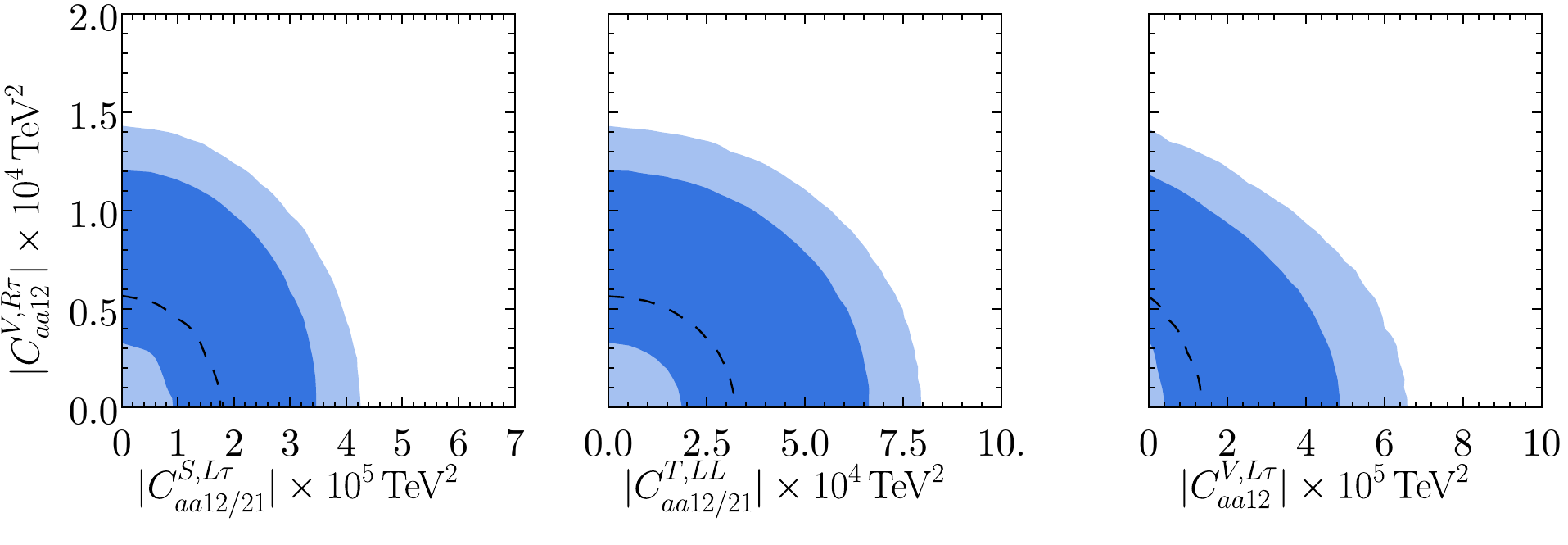}
  \includegraphics[scale=0.47]{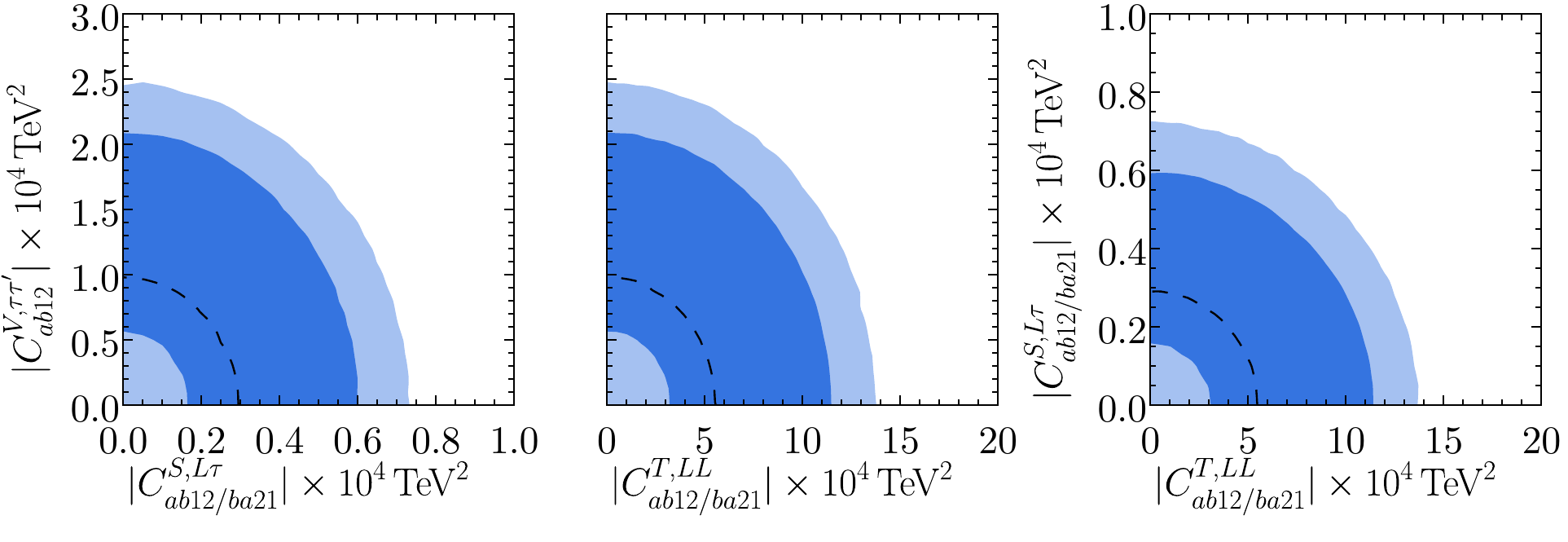}
  \caption{Allowed parameter space at $68\%$ CL (dark blue) and $90\%$ CL
  (light blue) in the plane of two NP Wilson coefficients 
  corresponding to various combinations of scenario (i), i.e. the case of three massless Dirac neutrinos. Note that here $a\neq b$.
  The dashed line corresponds to the future projection for the HIKE experiment at $90\%$ CL.
\label{fig:correlationsDirac}
}
\end{figure}
\begin{figure}[t]
  \centering
  {\large\bfseries Massless Majorana Neutrinos}\\
  \includegraphics[scale=0.47]{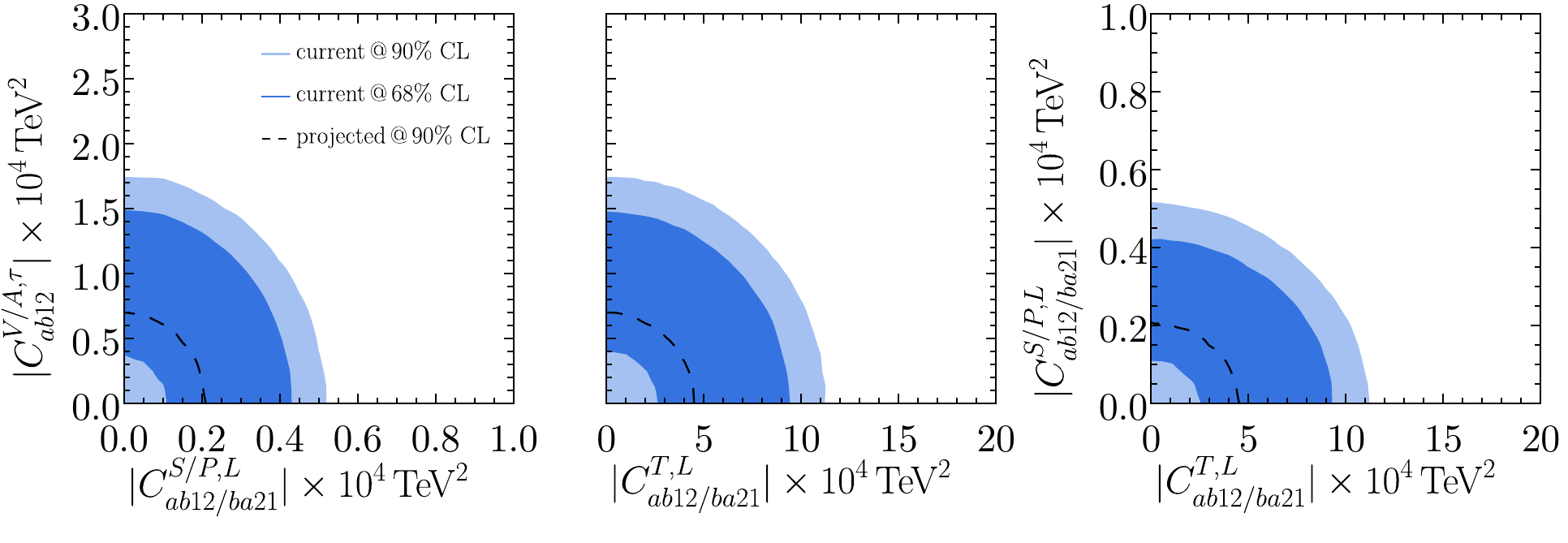}
  \includegraphics[scale=0.47]{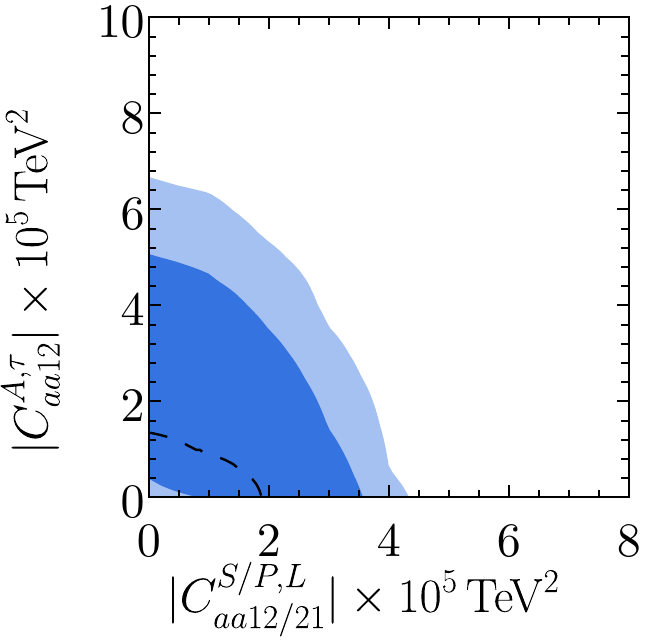}
  \caption{Allowed parameter space at $68\%$ CL (dark blue) and $90\%$ CL
    (light blue) in the plane of two NP Wilson coefficients 
    corresponding to various combinations of scenario (ii), i.e. the case of three massless Majorana neutrinos. Note that here $a\neq b$.
    The dashed line corresponds to the future projection for the HIKE experiment at $90\%$ CL.
  \label{fig:correlationsMajorana}
  }
\end{figure}

\subsection{Multi-operator correlations\label{sec:correlations}}

Next we study the correlations of probing different operators by considering the
effect of switching on pairs of different operators simultaneously. For
concreteness we cover (almost) all different non-trivial combinations for the case of
three massless Dirac or Majorana neutrinos, scenario (i) and (ii),
respectively. For these correlations it can be important to specify the phase we chose
for the NP Wilson coefficients, since relative phases can be relevant for interferences.
We assume there is no new CP-violating phases beyond the SM, i.e., all NP phases
are aligned to the SM one.
Additionally, in the discussion of neutrino-flavour diagonal operators, e.g., 
$O^{V,\,LL}_{aa12}$,
we consider the flavour-universal case in which we switch on all operators of the category 
simultaneously but with universal Wilson coefficients, e.g., 
$C^{V,\,LL}_{1112} =C^{V,\,LL}_{2212} =C^{V,\,LL}_{3312}$.
This additional assumption of flavour universality reduces the number of different combinations.

The results are shown in Figure~\ref{fig:correlationsDirac} and \ref{fig:correlationsMajorana}.
For each case we determine the allowed region at $68\%$ CL (dark blue region) 
and $90\,\%$ CL (light blue region) in the plane of two Wilson coefficients 
assuming that the NP scenario is realised in nature.
The dashed, black lines correspond to the  $90\%$ CL future projection.

\section{Discussion and conclusions\label{sec:conclusions}}

The golden, rare kaon decays $K\to\pi\nu\nu$ provide an excellent 
test of the Standard Model and are already placing stringent constraints on 
the parameter space of New Physics. 
The expected and planned experimental activities will test
modifications of the SM operators at energies of
$\mathcal{O}(300\,\text{TeV})$ and will also allow to extract
additional information from the invariant mass spectrum of the
neutrinos.
In the weak effective theory, scalar and tensor operators
induce a modified invisible mass distribution, c.f.~Figure~\ref{fig:distsandNA62}. 
We we have classified all independent Majorana and
Dirac operators that contribute to these FCNC
decays to neutrinos.
We have derived the sensitivity to each operator for the current
and expected future experimental situation including experimental and
theoretical uncertainties, see Tables~\ref{tab:MajSensitivities} and
\ref{tab:DiracSensitivities}.
The expected constraints on the NP scale range from $\mathcal{O}(40 \,\text{TeV})$ for
tensor operators to $\mathcal{O}(300 \, \text{TeV})$ for axial-vector  
Majorana and vector Dirac operators.
We have studied the impact of a massive sterile neutrino 
on the invisible mass spectrum and have discussed 
the correlation of various operators by analysing the 
allowed parameter space for pairs of different NP operators. 

Majorana and Dirac neutrinos, which can be generated from a lepton-number 
violating SMEFT or a lepton-number conserving $\nu$SMEFT theory
with additional right-handed neutrinos, have a completely different UV
origin. Therefore, the relevant weak effective theories also 
contain different number of independent parameters. This implies that 
concrete UV models could be tested from the data of the spectrum in future searches. 
The NP sensitivity of the invisible spectrum provides a strong motivation for the future Kaon program 
and we plan to study the discriminating power of the charged decay 
mode spectrum on different realisations of the SMEFT and $\nu$SMEFT 
in the future \cite{GMSSTinprep}.

\subsubsection*{Acknowledgments}
We thank Radoslav Marchevski for the discussion regarding the 
future prospects of the HIKE experiment.
Moldanazarova Ulserik was supported by the grant AP14972893 from the
Committee of Science of the Ministry of Science and Higher Education
of the Republic of Kazakhstan. The work of MG is partially supported
by the UK Science and Technology Facilities Council grant ST/X000699/1.
UM and MG thanks the particle theory group at the TU Dortmund for the hospitality
during the time in which this work was in part done.

\clearpage

\appendix
\section{Decay rates\label{app:DalitzDecayRates}}

The full expressions for the double differential decay widths
$\mathrm{d}^2\Gamma_{ab}/\mathrm{d}q^2\mathrm{d}|\vec{p}_\pi|$ and
$\mathrm{d}^2\Gamma_{ab}/\mathrm{d}q^2\mathrm{d}k^2$ for massive Majorana and
Dirac neutrinos are openly available in a git repository on
GitLab~\cite{gitlab}. Here we give expressions for the double differential
partial decay width in terms of the Lorentz invariant Dalitz variables $q^2 =
(p_K - p_\pi)^2$ and $k^2 = (p_\pi + p_{\nu,a})^2$ for massless
neutrinos. In the Majorana case we find
\begin{equation}
  \begin{aligned}
    \frac{\mathrm{d}^2 \Gamma_{ab}}{\mathrm{d}q^2\mathrm{d}k^2}\Bigg|_\text{Majorana} =&
    \frac{1}{512 \pi^3 m_K^3 (1 + \delta_{ab})} \Theta_{\text{L. inv.}}\left(q^2,k^2\right)\Bigl|_{m_a = m_b = 0}\Bigg[ \\
    & -4 \left|f^V(q^2)\right|^2 (k^4 - k^2 (m_K^2 + m_\pi^2 - q^2) + m_K^2 m_\pi^2) \\
    &\qquad \times \left(\left|C^{A,\,L, \,\text{SM}}_{ab 12} + C^{A,\,L}_{ab12} + C^{A,\,R}_{ab12}\right|^2 + \left|C^{V,\,L}_{ab12} + C^{V,\,R}_{ab12}\right|^2\right) \\
    & +q^2 \left|F^S(q^2)\right|^2 \Big(\left|C^{P,\,L}_{ab12} + C^{P,\,L*}_{ab21}\right|^2+\left|C^{S,\,L}_{ab12} + C^{S,\,L*}_{ab21}\right|^2 \Big) \\
    & +q^2 \left|F^T(q^2)\right|^2 (-2 k^2 + m_K^2 + m_\pi^2 - q^2)^2\Big(\left|C^{T,\,L}_{ab12} - C^{T,\,L*}_{ab21}\right|^2 \Big)\\
    & +2 q^2 (2 k^2 - m_K^2 - m_\pi^2 + q^2) \Big(    \\
    &\qquad  - F^S(q^2) F^T(q^2) \mathrm{Im}\Big(C^{S,\,L}_{ab12} C^{T,\,L}_{ab21} + C^{S,\,L}_{ab21} C^{T,\,L}_{ab12}\Big) \\
    &\qquad +F^S(q^2) \left(F^T(q^2)\right)^* \mathrm{Im}\Big(C^{S,\,L}_{ab21} C^{T,\,L*}_{ab21} + C^{S,\,L}_{ab12} C^{T,\,L*}_{ab12}\Big) \Big) \Bigg] \, ,
  \end{aligned}
\end{equation}
with the theta function in terms of $q^2$ and $k^2$ defined as
\begin{equation}\label{eq:theta-func}
  \begin{split}
    \Theta_{\text{L. inv.}}\left(q^2,k^2\right)\Bigl|_{m_a = m_b = 0} &= \theta(q^2) \theta\bigl((m_K - m_\pi)^2 - q^2\bigr)\\
      & \times \theta\left(k^2 - \frac{1}{2}\left(m_K^2 + m_\pi^2 - q^2 - \lambda^{1/2}_{K\pi q^2}\right)\right) \\
      & \times \theta\left(\frac{1}{2}\left(m_K^2 + m_\pi^2 - q^2 + \lambda^{1/2}_{K\pi q^2}\right)- k^2\right)\,.
  \end{split}
\end{equation}
For massless Dirac neutrinos we similarly find
\begin{equation}
  \begin{aligned}
    \frac{\mathrm{d}^2 \Gamma_{ab}}{\mathrm{d}q^2\mathrm{d}k^2} &= \frac{1}{256 \pi^3 m_K^3}\Theta_{\text{L. inv.}}\left(q^2,k^2\right)\Bigl|_{m_a = m_b = 0}\\
    & \times \Bigg( -(f^V(q^2))^2 (k^4 - k^2 (m_K^2 + m_\pi^2 - q^2) + m_K^2 m_\pi^2) \\
    & \qquad \times \left(\left|C^{V,\,LL,\,\text{SM}}_{ab 12} + C^{V,\,LL}_{ab 12} + C^{V,\,LR}_{ab 12}\right|^2 + \left|C^{V,\,RL}_{ab 12} + C^{V,\,RR}_{ab 12}\right|^2\right) \\
    & + \frac{q^2}{4} \left((F^S(q^2))^2 \left(\left|C^{S,\,LL}_{ab12}+C^{S,\,LR}_{ab12}\right|^2+\left|C^{S,\,LR}_{ba21}+C^{S,\,LL}_{ba 21}\right|^2\right)\right.\\
    & \qquad\left. +(F^T(q^2))^2 (-2 k^2+m_K^2+m_\pi^2-q^2)^2 \left(\left|C^{T,\,LL}_{ab12}\right|^2+\left|C^{T,\,LL}_{ba 21}\right|^2\right)\right) \\
    & - \frac{q^2}{2} (-2 k^2+m_K^2+m_\pi^2-q^2) \Big( \\
    & \qquad +  F^S(q^2) F^T(q^2) \mathrm{Im}\left(C^{T,\,LL}_{ab12} \left(C^{S,\,LL}_{ba21}+C^{S,\,LR}_{ba21}\right)\right)\\
    & \qquad + F^S(q^2) (F^T(q^2))^* \mathrm{Im}\left(C^{T,\,LL*}_{ba21} \left(C^{S,\,LL}_{ba21}+C^{S,\,LR}_{ba21}\right)\right)\\
    & \qquad + (F^S(q^2))^* F^T(q^2) \mathrm{Im}\left(C^{T,\,LL}_{ab12} \left(C^{S,\,LL}_{ab12}+C^{S,\,LR}_{ab12}\right)^*\right)\\
    & \qquad + (F^S(q^2))^* (F^T(q^2))^* \mathrm{Im}\left(C^{T,\,LL*}_{ba21} \left(C^{S,\,LL}_{ab12}+C^{S,\,LR}_{ab12}\right)^*\right)\Big) \Bigg) \, .
  \end{aligned}
\end{equation}
Note, that for massless neutrinos the interference term between
the scalar and tensor operators vanishes for both Majorana, and Dirac neutrinos, after integration over the
full $k^2$ phase space, which is why such terms do not appear in
Eqs.~\eqref{eq:dGmasslessMajorana} and \eqref{eq:dGmasslessDirac}.

\clearpage
\addcontentsline{toc}{section}{References}
\bibliographystyle{JHEP}
\bibliography{references}

\end{document}